\documentclass[10pt,journal,compsoc]{IEEEtran}
  \pdfoutput=1\relax                   
  \usepackage{graphicx}                
  \DeclareGraphicsExtensions{.pdf,.png,.jpg,.jpeg} 
  \usepackage{graphicx}                
  \DeclareGraphicsExtensions{.eps}     

\graphicspath{{figures/}{pictures/}{images/}{./}} 

\usepackage{microtype}                 
\PassOptionsToPackage{warn}{textcomp}  
\usepackage{textcomp}                  
\usepackage{mathptmx}                  
\usepackage{times}                     
\usepackage{cite}                      
\usepackage{tabu}                      
\usepackage{booktabs}                  
\usepackage{multirow}
\usepackage[table,xcdraw]{xcolor}
\usepackage{hyperref}
\usepackage[utf8]{inputenc}

\usepackage{ulem}
\usepackage{enumitem}
\usepackage{wrapfig}
\usepackage{caption}
\usepackage{etoolbox}
\usepackage{soul}
\usepackage{fontawesome}  
\usepackage{marvosym}     
\usepackage{float}
\usepackage{graphicx}

\newcommand{\rv}[1]{{\color{black}{#1}}}
\newcommand{\mr}[1]{{\color{black}{#1}}}
\newcommand{\bl}[1]{{\color{black}{#1}}}
\newcommand{\rr}[1]{{\color{black}{#1}}}

\begin{document}
\title{Leveraging Foundation Models for Crafting Narrative Visualization: A Survey }


\author{
Yi He*,
Ke Xu*,
Shixiong Cao,
Yang Shi,
Qing Chen,
and~Nan Cao
\IEEEcompsocitemizethanks{
\IEEEcompsocthanksitem Yi He, Shixiong Cao, Yang Shi, Qing Chen, and Nan Cao are with Intelligent Big Data Visualization Lab, Tongji University. \\Email: \{heyi\_11,caoshixiong,yangshi.idvx,qingchen,nan.cao\}@tongji.edu.cn. \\Nan Cao is the corresponding author. 
\IEEEcompsocthanksitem Ke Xu is with the School of Intelligence Science and Technology at Nanjing University. E-mail: lukexuke@gmail.com. 
\IEEEcompsocthanksitem * These authors contributed equally to this work.
}
}


\IEEEtitleabstractindextext{%


\begin{abstract}
Narrative visualization transforms data into engaging stories, making complex information accessible to a broad audience. \mr{Foundation models, with their advanced capabilities such as natural language processing, content generation, and multimodal integration, hold substantial potential for enriching narrative visualization. Recently, a collection of techniques have been introduced for crafting
narrative visualizations based on foundation models from different
aspects. We build our survey upon 66 papers to study how foundation models can progressively engage in this process and then propose a reference model categorizing the reviewed literature into four essential phases: Analysis, Narration, Visualization, and Interaction. Furthermore, we identify eight specific tasks (e.g. Insight Extraction and Authoring) where foundation models are applied across these stages to facilitate the creation of visual narratives. Detailed descriptions, related literature, and reflections are presented for each task. To make it a more impactful and informative experience for diverse readers, we discuss key research problems and provide the strengths and weaknesses in each task to guide people in identifying and seizing opportunities while navigating challenges in this field.}
\end{abstract}

}

\maketitle
\section{Introduction} 
\IEEEPARstart{N}arrative visualization, a powerful approach to transforming data into compelling visual data stories~\cite{hullman2011visualization}, is designed to be accessible and engaging for a wide audience. It is an effective combination of data narrative and information visualization, enabling data-driven stories that convey context, causality, and data insights.
Crafting a narrative visualization requires diverse creators' skills, including expertise in data analysis, design, storytelling, and data visualization\cite{shi2020calliope,sun2022erato}. These skills surpassed the capability of most existing automatic generation techniques. 

\mr{At the same time, foundation models such as Large Language Models~\bl{(LLMs)}, image generation models, and multimodal foundation models \cite{yuan2022roadmap} demonstrate remarkable capabilities~\cite{KHAN2023100026} that can be harnessed for generating visual narratives~\cite{dibia2023lida}. Unlike traditional AI and machine learning approaches, foundation models acquire extensive knowledge through pretraining, enabling rapid adaptation to novel tasks. Moreover, foundation models can process natural language inputs, lowering the threshold for users to perform complex tasks like data processing, design, or programming. 
Recently, many techniques have been introduced for crafting narrative visualizations based on foundation models from different aspects. However, a comprehensive review paper that systematically summarizes these techniques and highlights potential research directions is still missing. Therefore, we believe a dedicated review is desired to track these developments and analyze their implications for the crafting of narrative visualization.
}



\rv{This paper summarizes recent advances in using foundation models to craft a narrative visualization. A standard crafting reference model is summarized based on the literature review (Fig.~\ref{fig:pipeline}), which consists of four major steps : \textit{\textbf{Analysis}}, i.e.,  using foundation models to analyze the raw data to chose the relevant data content~\cite{chen2023beyond,hamalainen2023evaluating,tavast2022language,borisov2022language} or extract data insights~\cite{sun2022erato,lingo2023role,zhang2023concepteva,shrivastava2020iseql,9964397,10081462}; \textit{\textbf{Narration}}, i.e., using foundation models to construct the narrative logic and structure~\cite{gao2023chatgpt,gangal2022nareor,kiciman2023causal,10.1145/3544548.3580753,ghosh2023spatio} or generate narrative content such as title, caption, annotation, and script~\cite{tang2023vistext,shvetsova2023howtocaption,9973198,liew2022using,sultanum2023datatales,bromley2023difference}; \textit{\textbf{Visualization}}, i.e., using foundation models to generate charts with image embellishments and animations to facilitate information communication~\cite{lingo2023role,maddigan2023chat2vis,maddigan2023chat2vis2,dibia2023lida,wang2023llm4vis,10017653,city31527,kavaz2023chatbot,ghosh2023spatio,10.1145/3490100.3516473,schetinger2023n,10.1145/3544549.3583931}; \textit{\textbf{Interaction}}, i.e., using foundation models to support and implement complex user interactions such as nature language query and data exploration, which makes the narrative visualization more comprehensible, accessible, and controllable to its readers~\cite{liu2022deplot, zhou2023enhanced,10296056,rahman2023chartsumm,kantharaj2022chart,cheng2023chartreader,liu2022deplot,kantharaj-etal-2022-opencqa}. Specifically, the contributions of this paper are summarized as follows:}

\begin{itemize}
  \item \rv{We categorize existing literature on using foundation models for narrative visualization and introduce a reference model consisting of four stages and eight tasks, exploring how foundation models are used in supporting the generation of narrative visualization.}
  
  \item \rv{We made an in-depth discussion on the existing research limitations and future research opportunities to help inspire future research.}
\end{itemize}

The rest of the survey is structured as follows: \rv{In Section 2, we introduce the related work and the methodology for literature collection. In Section 3, we describe how to construct the reference model and provide a comprehensive definition of it.} In Sections 4, 5, 6, and 7, we present each stage along with the associated problems. Following that, in Section 8, we discuss the methods of evaluation. In Section \rv{9}, we delve into the discussion of research challenges and opportunities. Finally, we bring our survey to a conclusion in Section \rv{10}. In addition to the state-of-the-art survey, we developed an interactive browser to facilitate the exploration and presentation of the collected papers at \textbf{http://lm4vis.idvxlab.com/}.


\section{Related Survey and Methodology}
In this section, we first review the relevant surveys to highlight the value of our work. After that, we introduce the scope and methodology of our survey.

\subsection{Related Surveys}
\rv{For a long time, the process of creating visualization diagrams has been time-consuming and labor-intensive, requiring people to have skills in data analysis, design, and programming. With the rapid development of new techniques, artificial intelligence have been used to help with the generation of visualization diagrams.
These techniques have been well summarized in several relevant survey papers. 
For example, 
Wang et al.~\cite{wang2021survey} surveyed 88 ML4VIS papers, identifying seven main visualization processes where ML techniques can be applied, and aligning these with existing theoretical models in an ML4VIS pipeline. Wu et al. \cite{wu2021ai4vis} provided an overview of AI techniques applied to generate information visualization diagrams (AI4VIS). Di et al.~\cite{di2023doom} explores the challenges and opportunities of integrating generative models into visualization workflows. Yang et al. \cite{yang2024foundation} also broadly explored the integration of visualization and foundation models by respectively exploring visualization for foundation models (VIS4FM) and foundation models for visualization (FM4VIS). All these survey papers mainly focused on information visualization techniques, but none of them focused on visual narrative. Recently, Chen et al.~\cite{chen2023does} made a comprehensive survey on visualization tools to explore how automation engages in visualization design and narrative processes. They summarized six narrative visualization genres and categorized them by intelligence and automation levels. This work is the most relevant to our survey, however, the authors only focused on tools instead of foundation models and detailed techniques.}

Compared to the above survey papers,  we would like to explore how foundation models are used for generating narrative visualizations from different aspects, including analysis, narration, visualization, and interaction. 
\mr{Especially, a foundation model is any model trained on broad data, typically using self-supervision at scale, which can be adapted (e.g., fine-tuned) to a wide range of downstream tasks. Current examples include BERT, GPT, and CLIP~\cite{yuan2022roadmap,bommasani2021opportunities }.} In this paper, we focus on models with over 100 million parameters. 

\subsection{Literature Selection}

We collect papers from a set of relevant journals and conference proceedings using both the \textbf{search-driven} and \textbf{reference-driven} methods. In particular, \rv{for the search-driven selection, we conducted four rounds of paper selection:}
\begin{table}[htbp]
  \centering
  \caption{Relevant Venues}
    \begin{tabular}{lp{20em}}
    \toprule
    \textbf{VIS} & VIS(infoVIS VAST), PacificVIS\newline{}TVCG, EuroVIS(Computer graphics forum) \\
    \midrule
    \textbf{HCI} & \multicolumn{1}{l}{CHI, TOCHI, UIST, IUI, CSCW, IJHCS} \\
    \midrule
    \textbf{AI} & AAAI, IJCAI, TPAMI, AI\newline{}ICML, NeurIPS, ICLR, JMLR(Machine Learning)\newline{}CVPR, IJCV, ICCV(Computer Vision)\newline{}ACL, EMNLP(Natural Language Processing)\newline{}TIST, TiiS \\
    \bottomrule
    \end{tabular}%
  \label{tab:venues}%
\end{table}%

\rv{\textbf{Identification:}} we select relevant papers from the top venues in the field of artificial intelligence, data visualization, and human-computer interaction as summarized in Table~\ref{tab:venues}. We specifically select papers containing keywords related to foundation models (e.g., `generative AI,' `large language models,' `pre-trained models,' `foundation models,' `image-generation models,' `GPT,' `AIGC') and narrative visualization (e.g., `visualization,' `infographic,' `data videos,' `narrative,' `data comics,' `annotated chart,' `slideshow,' `data story'). As a result, 2417 papers were selected.
\begin{figure*}[htbp]
    \centering
    \includegraphics[width=1\textwidth]{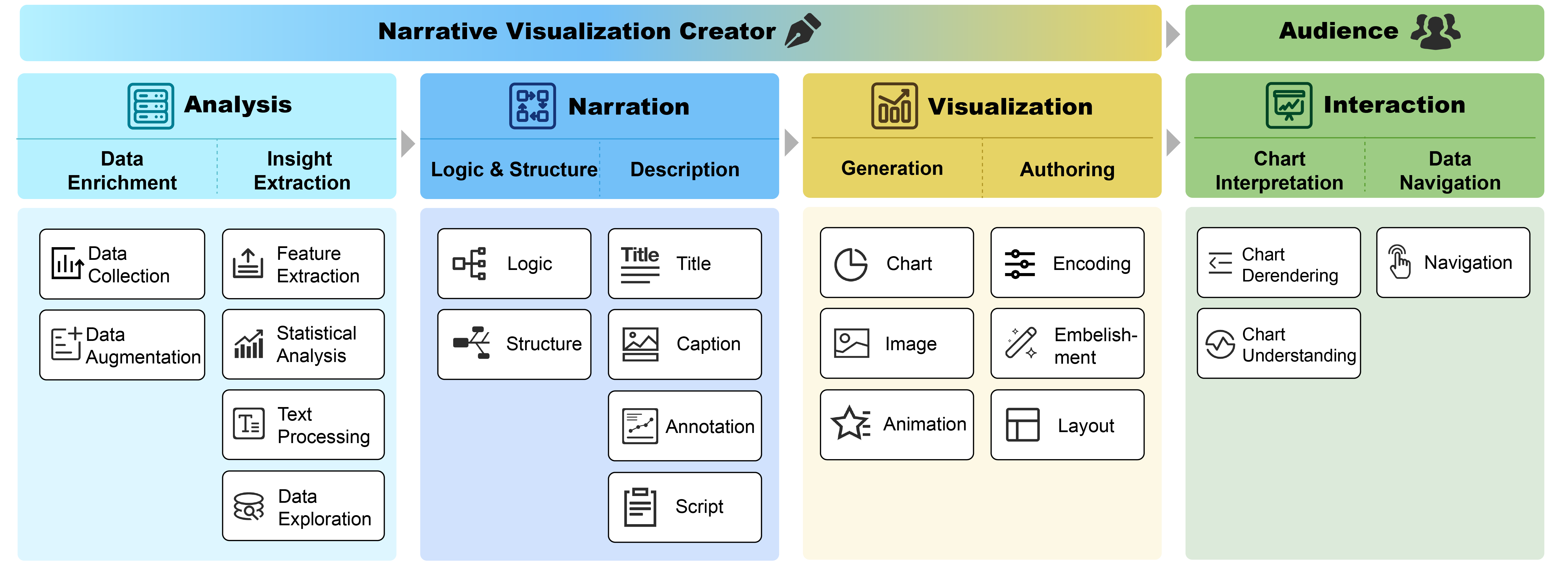}
    \caption{A Reference Model for Creating Narrative Visualizations Based on Foundation Models}
    \label{fig:pipeline}
\end{figure*}
\rv{\textbf{Code Screening:}} we used a program for preliminary screening and compiled a keyword list that included terms related to narrative visualization and foundation models. We matched these keywords against the titles and abstracts of the initially selected papers, including only those that encompassed both aspects in our selection.  As a result, only 361 papers are retained.
\begin{figure}[htbp]
    \centering
    \includegraphics[width=0.5\textwidth]{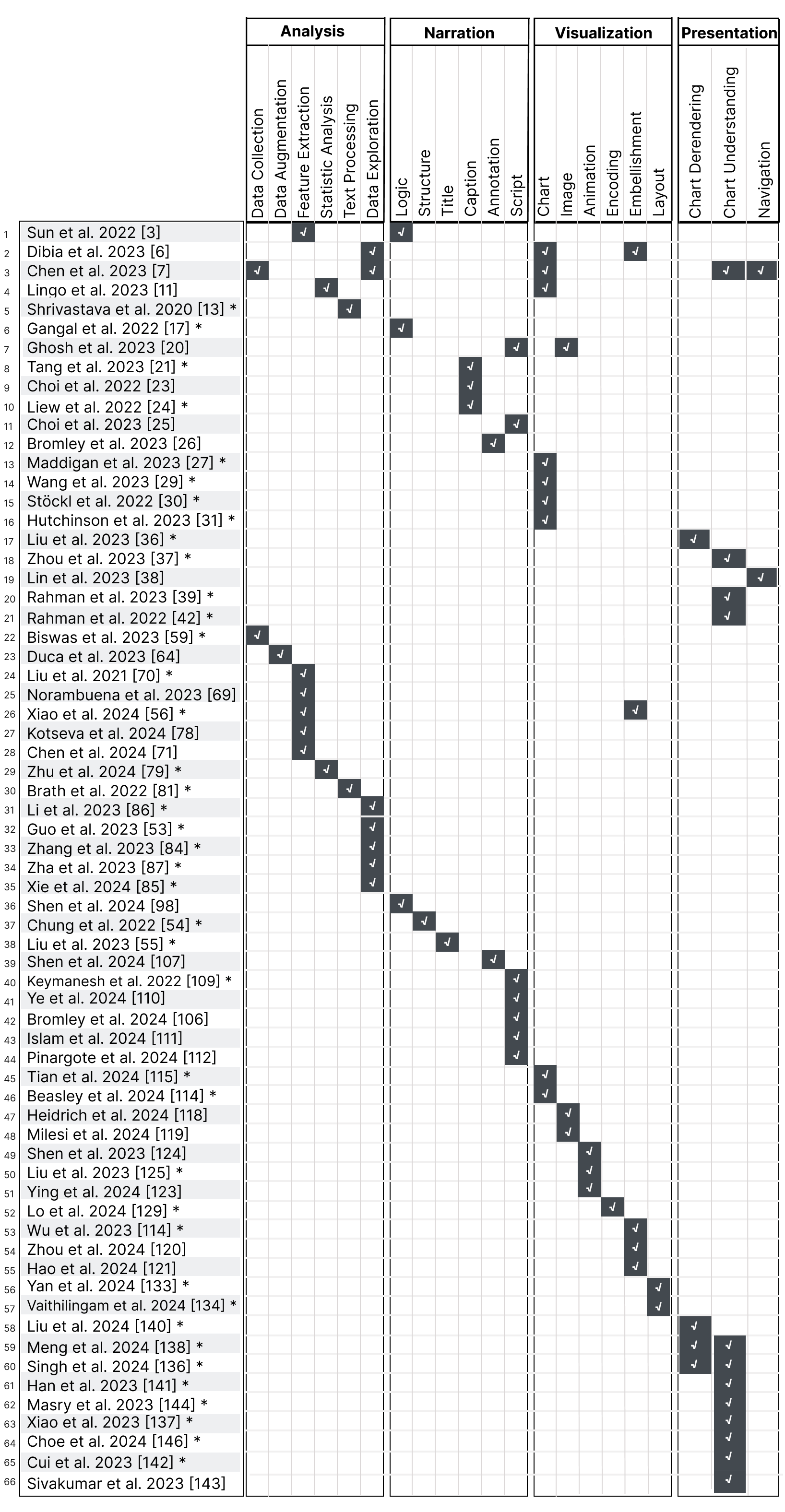}
    \vspace{-4mm}
    \caption{\mr{An overview of the papers related to the reference model and their associated code. Papers marked with (*) represent tools that can be used in creating narrative visualizations, even if the primary focus of the research is not specifically on narrative visualization. \rr{A clearer and more detailed collection of the papers can be found at http://lm4vis.idvxlab.com/.} }}
    \vspace{-6mm}
    \label{fig:paper}
    
\end{figure}
\rv{\textbf{Manual Screening:}} 
\rv{we meticulously reviewed the abstracts of the remaining articles to ensure their alignment with our research scope.} This thorough assessment allowed us to refine our selection, resulting in 121 articles being included in our research corpus.

\rv{\textbf{Eligibility Assessment:} finally, we checked the eligibility of each paper based on the following exclusion criteria : (1) articles using foundation models with fewer than 100 million parameters were excluded;  \bl{(2) articles that are irrelevant to the process of crafting narrative visualizations were excluded. }As a result, 121 papers were filtered out and we conducted a detailed full-text review of these papers. This meticulous review process involved carefully examining each article to ensure it met all inclusion criteria. We paid particular attention to the methodologies, results, and discussions presented in each article to confirm their alignment with the scope of our study. Finally, only \bl{55} of the most relevant papers were selected in our review corpus. }

\rv{In the reference-driven selection, we went through all the papers cited in the above \bl{55} papers.} Some important papers (i.e., highly relevant and highly cited) from journals or conference proceedings outside our searching scope were also included in our review corpus. Finally, we have compiled a total of \bl{66} most relevant papers. The papers in our corpus are meticulously cataloged in Fig. \ref{fig:paper}.  

\subsection{Literature Review}
\bl{The paper analysis was conducted in three phases. In the first phase, we extracted a concise description for each paper, including the foundation model mentioned (e.g., GPT, BERT, etc.), the type of foundation model, the scale of training parameters used, the specific visualization problem addressed by the model, the context or scenario where the model was applied, and a summary of the paper's main theme or contribution.
In the second phase, four authors summarized and merged similar items of the encoding above, enabling us to identify common patterns and determine categories. During weekly discussions, we iteratively adapted, split, and refined the codes multiple times until no further disagreements arose. We then classified the literature into four stages and eight main tasks., with eachtask further divided into detailed subtasks specifying the specific activities involved.
In the third phase, each author was responsible for classifying papers within a specific stage until ultimately completing the paper selection.
}

\section{Reference Model}\label{sec:reference-model}
\mr{To understand how foundation models support the generation of narrative visualizations, it is essential to analyze the creation process and identify the key techniques at each generation stage where foundation models can contribute effectively.

Several researchers have attempted to outline the process of creating narrative visualization. For instance, Lee et al.\cite{7274435} identified three primary phases in crafting data stories: exploring data, making a story, and telling a story. Segel et al. summarized three key elements of narrative visualization: narrative genres, narrative structure, and visual narrative \cite{segel2010narrative}. Li et al.\cite{10.1145/3613904.3642726} also investigated a framework encompassing analysis, planning, implementation, and communication. Amini et al. \cite{10.1145/2702123.2702431}, through observations of how people create data videos, proposed a process involving data reading and interpretation, data selection, narrative crafting, and viewer engagement.
However, these summaries still have some limitations. Specifically, they tend to generalize the process of story design and lack detailed guidance on specific design actions that designers can undertake during the creation process.

To provide a more comprehensive understanding of how foundation models contribute to the creation of narrative visualization, we have synthesized and refined previous research findings. As illustrated in Fig.~\ref{fig:pipeline}, we introduce a reference model for creating narrative visualizations based on foundation models.} The reference model consists of four major stages, including analysis, narration, visualization, and interactions. In particular, the first three stages focus on the tasks that an analyzer needs to perform when creating a narrative visualization. The last stage focuses on how an audience interacts with an existing visualization to gain more insights.

\begin{wrapfigure}[3]{l}{0.045\textwidth}
    \vspace{-0.4cm}
    \includegraphics[width=0.06\textwidth]{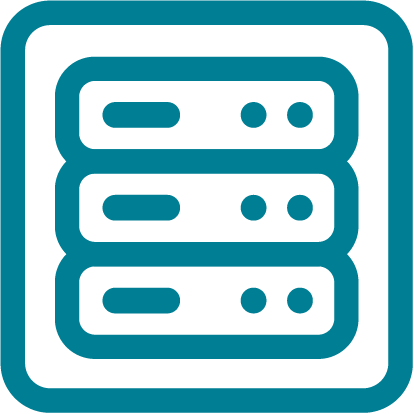}
\end{wrapfigure}
\noindent 
\rv{
\textbf{Analysis:}
the analysis stage deals with the input data and extracts meaningful data insight for narration regarding the story creators' requirements~\cite {wang2021survey}.  Here, foundation models are often used to help with data processing and uncover data insights. In particular, they are frequently used in two major tasks: (1) data enrichment and (2) insight extraction. \textbf{\textit{Data enrichment}} is to augment the raw data to improve its quality and usability. Here, foundation models are used either to find relevant datasets~\cite{chen2023beyond} or fix data errors such as eliminating outliers or imputing missing data~\cite{borisov2022language}.
\textbf{\textit{Insight extraction}} is to use foundation models to automatically analyze the raw data to extract useful knowledge regarding a narration topic~\cite{guo2023urania}.}

\begin{wrapfigure}[3]{l}{0.045\textwidth}
    \vspace{-0.4cm}
    \includegraphics[width=0.06\textwidth]{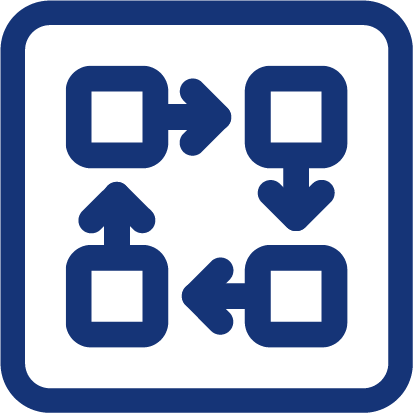}
\end{wrapfigure}
\noindent 
\rv{\textbf{Narration:}
the narration stage involves arranging a series of data insights logically and structurally to create a coherent and engaging story. Here, foundation models are often used to identify logical relationships between data insights and organize them. In particular, they are frequently used in two major tasks: (1) narrative logic and structure construction and (2) narrative content generation.
\textbf{\textit{Narrative logic and structure construction}} is to deliberately guide the audience's attention in a specific sequence, helping them maintain a sense of direction throughout the story. Here, foundation models can identify relationships between data facts and adjust the narrative order \cite{chung2022talebrush}.
\textbf{\textit{Narrative content generation}} is to use foundation models to generate summarizing or descriptive text\cite{10168257}.}

\begin{wrapfigure}[3]{l}{0.045\textwidth}
\vspace{-0.4cm}
    \includegraphics[width=0.06\textwidth]{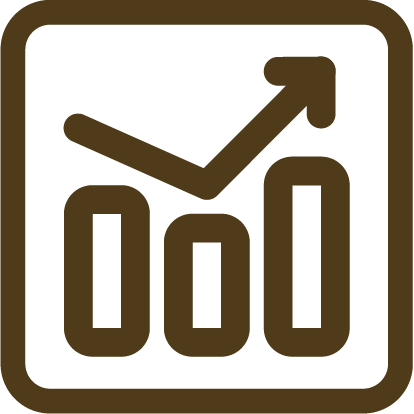}
\end{wrapfigure} 
\noindent 
\rv{\textbf{Visualization:}
the visualization stage transforms the prepared data insights into visual representations for the audience. 
Here, foundation models are often used to help users automatically generate accurate and visually appealing narrative visualizations. In particular, foundation models are primarily used to accomplish two main tasks: (1) visualization generation and (2) visual content embellishment.
\textbf{\textit{Visualization generation}} is to interpret data insights and select suitable methods for visual representation. Here, foundation models are often used to recommend the most suitable chart types and generate the necessary code to create charts\cite{dibia2023lida}.
\textbf{\textit{Visual content embellishment}} is to use foundation models to generate stylized visualization using the text-conditioned
image-to-image generation capabilities\cite{dibia2023lida,xiao2023let}.}


\begin{wrapfigure}[3]{l}{0.045\textwidth}
    \vspace{-0.4cm}
    \includegraphics[width=0.06\textwidth]{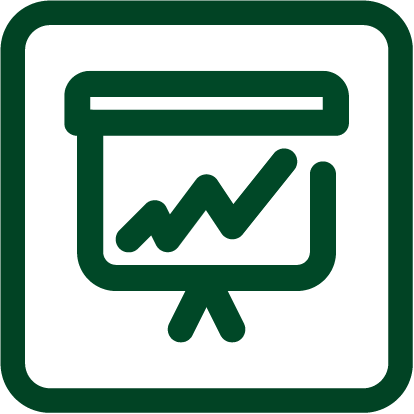}
\end{wrapfigure}
\noindent 
\rv{\textbf{Interaction:} 
the interaction stage deals with how users read and interact with the visualization to extract valuable information\cite{wang2021survey}. Here, foundation models can help the audience understand the narrative visualization more thoroughly and explore further. In particular, foundation models are primarily used to accomplish two main tasks: (1) facilitating natural language interactions and (2) enhancing interactive exploration. \textbf{\textit{Facilitating natural language interactions}} is to use foundation models to communicate with the audience using natural language, improving the effectiveness of information communication\cite{kantharaj-etal-2022-opencqa}. \textbf{\textit{Enhancing interactive exploration}} is to use foundation models to generate code to interact with visualizations by triggering corresponding events in the underlying JS code, such as switching data mappings, zooming in, and zooming out\cite{chen2023beyond}.}

\rv{In conclusion, foundation models play different roles in various stages to help users complete the process of creating narrative visualizations. These models provide robust support, allowing users to focus on the narrative content and the knowledge they want to convey rather than technical details. In the following sections, 4 through 7, we will provide a detailed overview of the tasks foundation models accomplish. At the end of each subsection, we explore the existing challenges and future research directions.}

\section{Analysis}
\rv{To tell a data story or craft a narrative visualization, the first thing an author should handle is to collect and analyze the data. We reviewed 19 papers related to the data analysis stage, which can be broadly classified into two major categories based on the role of foundation models: (1) data enrichment and (2) insight extraction, which will be discussed in this section.}

\subsection{Data Enrichment}

\begin{wrapfigure}[3]{l}{0.045\textwidth}
    \vspace{-0.4cm}
    \includegraphics[width=0.06\textwidth]{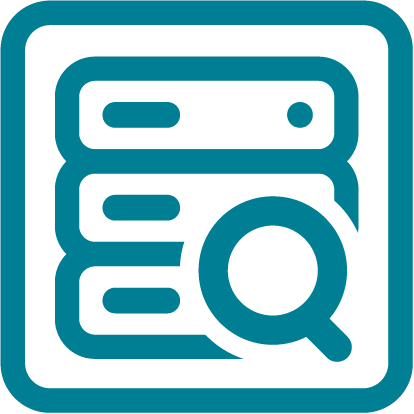}
\end{wrapfigure}

Data is not a natural resource~\cite{di2023doom}. \rv{When crafting a narrative visualization, users usually need to gather datasets from various sources, which is usually time-consuming and labor-intensive, especially for large-scale datasets. In addition, the collected raw data often contain missing values, inconsistencies, or even errors, which can greatly impact the accuracy of subsequent analyses. To address these challenges, recent research has focused on two main aspects: data collection and data augmentation based on foundation models.}

\textbf{Data Collection} involves integrating multiple data sources\cite{deng2009imagenet}. Foundation models, with their information retrieval capabilities, can align users' narrative goals and requirements with various data sources and effectively identify the most relevant content~\cite{kalyan2023survey}. For example, 
Chen et al.\cite{chen2023beyond} demonstrated the utility of GPT-4 in suggesting relevant website links and providing access to downloadable datasets, thus facilitating the collection of diverse and pertinent data. Similarly, Biswas et al.\cite{biswas2023function} showed that ChatGPT could collect and organize data from social media platforms. Both studies highlight the ability of foundation models to access a wide range of data sources and aggregate relevant information.

\textbf{Data Augmentation} is the process of augmenting the existing datasets to make them more complete and comprehensive~\cite{losada2022day,10.1145/3543829.3544529, 10.1145/3589335.3651929}. Here, \rv{Foundation models are frequently used to augment existing datasets using Retrieval-Augmented Generation (RAG) technology. RAG is a natural language processing technique that combines information retrieval and generation models. RAG's main idea is to enhance the content generation process by retrieving relevant information from external knowledge bases\cite{NEURIPS2020_6b493230}. Using RAG, foundation models can integrate relevant information according to specific domain requirements for users, thereby providing richer context and background information for narrative visualization. For example, Duca et al. \cite{duca2024using} proposed a method using RAG technology to generate context for data-driven stories. First, documents related to specific topics were input into a vector database for indexing to facilitate efficient retrieval. Then, in the retrieval step, relevant information was queried from external knowledge bases. In the generation step, this information was integrated into an LLM, which generated accurate and detailed textual content based on the retrieved knowledge. This approach supplemented data narratives with rich background knowledge and contextual information, achieving the goal of data augmentation.}

\rv{\textbf{Reflection.} Current research shows that foundation models have substantial potential to automate the laborious process of data collection and augmentation\cite{kalyan2023survey}. Data collection involves using foundation models' information retrieval capabilities to identify and aggregate relevant datasets from diverse sources. Data augmentation is enhanced by foundation models through the use of RAG, which integrates external knowledge to enrich and supplement existing datasets.}

\rv{However, the generated content may contain errors\cite{rawte2023survey}. To prevent foundation models from generating incorrect information, enhancing the model's ability to recognize and acknowledge its knowledge gaps is essential\cite{feng2024don}. Therefore, future research should focus either on \uline{developing algorithms to help 
foundation models identify their limitations and avoid generating inaccurate content}~\cite{10.1145/3643834.3660685}, or \uline{integrating real-time human feedback to help refine the outputs generated by foundation models} ~\cite{xiao2023let,10.1145/3643834.3660685}. Research should also explore methods for displaying the basis and context of AI-generated content, making it more trustworthy.}

\subsection{Insight Extraction}

\begin{wrapfigure}[3]{l}{0.045\textwidth}
    \vspace{-0.4cm}
    \includegraphics[width=0.06\textwidth]{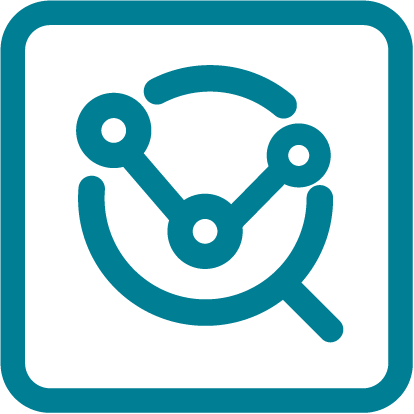}
\end{wrapfigure}

Insight extraction involves performing in-depth analysis to identify crucial insights and patterns in the data~\cite{7192662}. \rv{Foundation models, with their advanced analytical capabilities, enable users to go beyond superficial data analysis, allowing for deeper exploration and discovery of hidden value within the data\cite{kalyan2023survey}. In particular, recent research in this direction has focused on four main aspects: feature embedding, \rr{statistical analysis}, text processing, and data exploration.}

\begin{figure*}[htbp]
    \centering
    \includegraphics[width=1\textwidth]{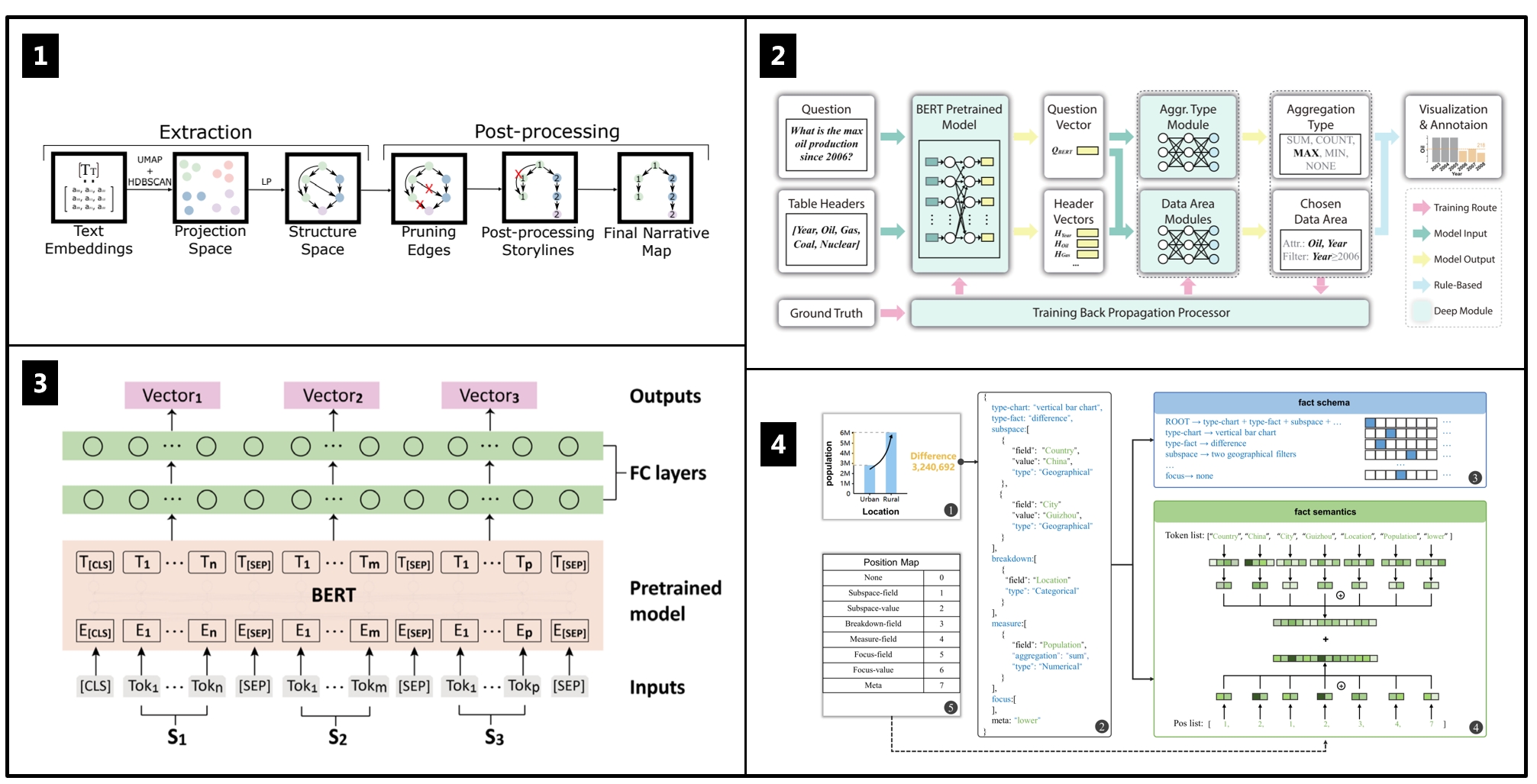}
    \caption{Selected examples of feature embedding: (1) Use text embedding technology to capture the deep semantic information of text, supporting the construction of graph-based narrative visualizations\cite{10.1145/3581641.3584076}.  (2) ADVISor: Use BERT to convert the natural language in table headers and questions to vectors presenting the
semantic meaning\cite{liu2021advisor}.  (3) Erato: schematic diagrams of the fact embedding model\cite{sun2022erato}. (4) Chart2Vec: learn a universal embedding of visualizations with context-aware information \cite{10485458}.}
    \vspace{-2mm}
    \label{embedding}
\end{figure*}
\rv{\textbf{Feature Embedding} converts unstructured data, like text \cite{cao2016overview}, images \cite{lotfi2023storytelling}, and structured data, like data facts~\cite{sun2022erato,10485458}, into semantic feature vectors~\cite{wen2021time,yang2024foundation}. Foundation models such as BERT \cite{devlin2018bert} and ELMo \cite{sarzynska2021detecting} transform textual data into concise, low-dimensional vector spaces.  In narrative visualization, these feature vectors can be used to identify patterns, trends, and connections within the data, which in turn enhances the storytelling aspect by providing deeper insights and more coherent narratives\cite{10.1145/3581641.3584076, xiao2023let, maharana2022storydall,liu2021advisor,kotseva2023trend}. For example, 
Norambuena et al.\cite{10.1145/3581641.3584076} used text embedding to capture deep semantic information from text, supporting the construction of graph-based narrative visualizations. The paper took news articles as input data, where each article was considered an event. By using the all-MiniLM-L6-v2 model to generate text embeddings, and combining UMAP algorithm for dimensionality reduction and HDBSCAN algorithm for clustering, it calculated the coherence values between events to construct a high-level discrete structure narrative map. This framework supported analysts in incrementally refining the narrative model through semantic interactions, leading to a better understanding of the relationships between news events, as shown in Fig. \ref{embedding} (1). These vector were used to generate the projection space and compute coherence.
Another example is ADVISor\cite{liu2021advisor}(Fig. \ref{embedding} (2)), which automatically generates annotated charts to answer natural language questions about tabular data based on BERT~\cite{devlin2018bert}. Here, BERT encodes natural language questions and table headers into latent vectors. These vectors are then used to extract relevant data regions from the table and identify potential data aggregation methods. Based on this information, charts are generated.

Despite the above techniques, recent research also focuses on how to fine-tune foundation models to design effective embedding methods that map data facts appropriately in a vector space that reflects the logical and semantic relationships of data facts. For example, Erato\cite{sun2022erato} fine-tuned a BERT to convert data facts into vectors with the distances between two vectors indicating logical relationships between data facts. In particular, two fully connected layers were added on top of BERT (Fig. \ref{embedding} (3)). The input format converted each data fact into a tokenized string, capturing both structural and semantic information. The model was fine-tuned on a dataset of manually designed data stories, where each training sample was a trigram of data facts capturing a logical relationship in the story. Another notable work is Chart2Vec\cite{10485458}. It proposed a method to create an embedding for narrative visualizations that incorporates context-aware information (Fig. \ref{embedding} (4)). In this work, authors collected a dataset of multi-view visualizations, including data stories from Calliope\cite{shi2020calliope} and dashboards from Tableau Public, and transformed these visualizations into data facts similar to Erato. The structural information was encoded using a one-hot vector representation of the parse tree derived from a context-free grammar, while the semantic information was encoded using pre-trained Word2Vec embedding for the textual content. These features were then fused, including positional encoding to preserve the connection between structural and semantic data, creating a final vector representation for each chart. The chart embedding results support various downstream tasks such as visualization recommendation and data story generation. \bl{Compared to general visualization, feature embedding in narrative visualization focuses more on the relationships between data facts and the contextual semantics. This emphasis helps in crafting a coherent and logical story.}


 }

\rr{\textbf{Statistical Analysis}} provides an overview of a dataset's central tendencies, dispersion, and overall distribution\cite{lingo2023role}. These statistics offer a preliminary understanding of the data, highlighting key attributes such as the mean, median, range, and standard deviation, and laying the groundwork for more in-depth data insight.

\rv{Foundation models excel in generating code to meet users' computational needs. For example, Zhu et al.~\cite{zhu2024large} used pre-designed templates to generate statistical questions, which were refined by GPT-3.5-Turbo. The model preprocessed the tabular data, such as cleaning and standardizing formats, and generated SQL queries or statistical code to perform the actual calculations. The results were then validated and interpreted to ensure accuracy and reliability. Similarly, Lingo et al.~\cite{lingo2023role} demonstrated that by adjusting prompts, ChatGPT could perform summary statistics, data grouping, hypothesis testing, and other operations on tabular data by generating code to provide real-time computational assistance.}
\bl{Both studies validate the capability of foundation models to perform statistical computations on tabular data, lowering the learning curve and barriers compared to traditional complex data analysis methods. However, for narrative visualization, it is essential not only to compute the statistical data but also to highlight the exciting results, such as outliers and change points, which can drive the story. These statistical data can serve as anchor points, drawing the reader's attention and encouraging further exploration.
}

\textbf{Text Summarization} \rv{\bl{extracts key information and insights from large volumes of textual data, condensing it to provide a clear summary that helps users quickly grasp the story behind the data\cite{santana2023survey,brath2022summarizing}.} \bl{Compared to traditional topic modeling methods in NLP like }NewsViews\cite{10.1145/2556288.2557228}, foundation models help users quickly understand the semantics of input text, offering preliminary exploration and visualization suggestions. Textual data, such as user feedback, comments, and subjective evaluations, are important sources for narrative visualization. However, they often present challenges for users in quickly organizing and extracting key insights\cite{brath2022summarizing}.
 However, emerging NLP techniques of foundation models like document summarization addressed these challenges. 
 For example, iSeqL\cite{shrivastava2020iseql} used ELMo to generate context-aware word embedding, allowing users to annotate text data interactively and update model predictions in real-time. Through iterative optimization and visualization tools, users could continuously adjust annotations and model parameters, gaining deeper insights into textual data distribution and model performance, thereby effectively conducting exploratory data analysis.}

\textbf{Data Exploration}
is to extract knowledge from data. Query-based data exploration is a simple example. This approach involves extracting insights through Question-Answering (QA) interactions~\cite{lei2018ontology}. Foundation models have demonstrated remarkable proficiency in interacting with user-generated prompts and questions to facilitate data exploration. These models, equipped with advanced natural language processing and machine learning capabilities, can interpret and respond to a wide range of user inquiries, allowing users to delve into datasets and uncover insights in an intuitive and user-friendly manner\cite{guo2023urania,zhang2023data,xie2024waitgpt}, improving human-computer interaction by moving away from complex commands and facilitating a more intuitive and user-friendly mode of interaction\cite{kalyan2023survey}. For example, \rv{Sheet Copilot~\cite{li2023sheetcopilot} could autonomously manage, process, analyze, predict, and visualize data. When a request was received, they transformed raw data into informative results that best aligned with the user's intent. Furthermore, TableGPT\cite{zha2023tablegpt} leveraged foundation models for comprehensive table understanding by encoding entire tables into global representation vectors, enabling complex data manipulation and analysis. It utilized a Chain-of-Command approach to break down complex tasks into simpler steps, executing them sequentially. Additionally, TableGPT enhanced its performance in specific domains through domain-specific fine-tuning and supports private deployment to ensure data privacy. TableGPT excels by specifically optimizing and fine-tuning LLMs for table-related tasks, enabling comprehensive table understanding and utilizing a chain-of-command approach to break down complex tasks into simple steps. Additionally, TableGPT supports domain-specific fine-tuning and private deployment, ensuring data privacy and security, making it superior to SheetCopilot in handling complex table tasks and sensitive data.}

\textbf{Reflection.} 
Feature embedding with foundation models transforms unstructured and structured data into semantic vectors. \rr{Statistical analysis} using foundation models generates and validates computations on tabular data. Text summarization through foundation models efficiently condenses and interprets textual data, facilitating exploration and enhancing the integration of qualitative insights into visualizations. Data exploration with foundation models is enhanced by their ability to interpret and respond to user-generated queries, enabling intuitive and interactive knowledge extraction from datasets. 

However, most research focuses primarily on the extraction of textual features\cite{10.1145/3581641.3584076,kotseva2023trend}. As the demand for more complex narrative visualizations grows, particularly in data videos, there is a need for effective integration and alignment of multimodal information, including visuals, audio, and text\cite{chen2023does}. Furthermore, studies have shown that integrating multimodal information can help users better understand and remember visualizations\cite{Errey2023}. Therefore, exploring the potential of large models in processing multimodal data becomes crucial for enhancing the quality of narrative visualizations.
Future research should focus on \uline{the integration of multimodal information, exploring the application of foundation models in multimodal data processing and how to achieve semantic alignment across different modalities.}  This will aid in processing multimodal data, providing more material for narrative visualizations, and designing semantically consistent multimodal narrative visualizations.
\mr{Furturemore, \uline{future research should investigate Chain-of-Thought strategies and systematically designed prompt engineering}\cite{gao2023chatgpt,10.1145/3491102.3517582}. These approaches can enhance the interpretability and accuracy of foundation models, while also clearly defining model tasks to prevent aimless exploration and improve the efficiency of narrative visualization creation.}

\section{Narration}
We reviewed 17 papers related to the narration stage in our reference model. These papers are broadly classified into two major categories based on the roles played by foundation models, i.e., (1) logic generation and (2) structure generation, which will be discussed in this section.
\subsection{Logic \& Structure}

\begin{wrapfigure}[3]{l}{0.045\textwidth}
    \vspace{-0.4cm}
    \includegraphics[width=0.06\textwidth]{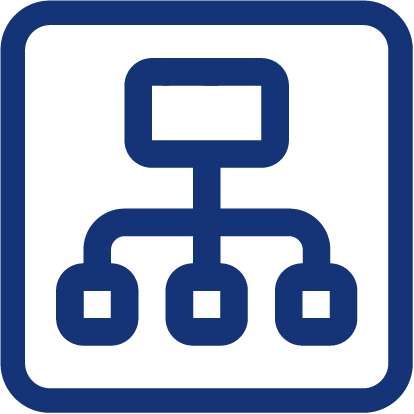}
\end{wrapfigure}

Logic and structure are the backbones of narrative visualization~\cite{mckenna2017visual}. An effective structure with a coherent logic will directly help with the understanding and memorization~\cite{OBIE2019113}.  In a data story, editors usually interlink data facts that are logically relevant to craft a storyline following a narrative structure~\cite{pietschmann2014transmedia}. However, analysis to detect data logic and fit the content into a narrative structure usually introduces technique barriers that are difficult for users with little data analysis knowledge to achieve. Foundation models, especially the pre-trained language models such as GPT~\cite{radford2018improving} and Bert~\cite{devlin2018bert}, have a great ability to help with logic reasoning in nature languages, which has also been used to build data logic~\cite{sun2022erato,10485458} and organize structural data narrations~\cite{chung2022talebrush}, recently.

\vspace{0.1cm}
\rv{\textbf{Logic} refers to the relationship between data facts\cite{10.1162/0891201054223977}. A disordered narrative sequence can confuse the audience due to the lack of logical dependence between the elements~\cite{wang2018narvis}. Based on large amounts of training data, foundation models can identify and capture logical relationships within data facts, which helps users automatically arrange the positioning of events when creating narrative visualizations.
For example, Erato \cite{sun2022erato} and Chart2Vec\cite{10485458}, as introduced in Section 4, converted data facts into vectors whose similarities directly indicated not only the semantic relationships but also the underlying logical connections between the corresponding facts. These approaches allow foundation models to effectively identify and represent complex data relationships within their respective contexts. Additionally, foundation models can also function as intelligent agents\cite{huang-chang-2023-towards}. By leveraging their inherent capabilities in reasoning and contextual understanding, these models can be guided to perceive and respond to user input, exhibiting goal-oriented behavior~\cite{xi2023rise}. For example, Shen et al.~\cite{shen2024data}
 proposed Data Director to automatically create data videos. In the Data Director system, GPT-4 was utilized as an intelligent agent to generate narrative text that established logical connections between data facts. Specifically, the system placed the GPT-4 in the role of a data analyst. By following a chain-of-thought approach, carefully designed prompts guide the GPT-4 to generate narrative text that not only connects insights into a coherent story but also ensures that the logic is compelling. The prompts explicitly instruct GPT-4 to avoid merely enumerating insights, thereby encouraging the creation of a more integrated and logically sound narrative.}

\vspace{0.1cm}
\textbf{Structure}, compared to logic, is a higher-level concept that determines how the entire narrative content is organized. There are many types of structures~\cite{10.1145/3411764.3445344,  8017584}. One famous example is known as Freytag’s Pyramid Structure~\cite{yang2021design}, which organizes the narrative content into four different stages, including initializing, rising, peaking, and releasing.

\rv{Foundation models have been used to help users interactively create coherent narratives that conform to a specific structure.
Some advanced techniques have been developed. In particular, 
instead of generating narrations based on a fixed structure, TaleBrush~\cite{chung2022talebrush} allows users to interactively create a structure by simply sketching a horizontal curve on canvas. GPT is then used to generate the narrative content based on this curve with the changing slopes in the curve indicating the positive or negative change of the generated storyline. In this way, by using this tool, users can create data stories with multiple climaxes and turning points in the plot.


Despite the interactive structure adjustment,  narrative structures can also be automatically adjusted by fine-tuning the foundation models. For example, NAREOR\cite{gangal2022nareor} fine-tuned foundation models using two innovative training methods to control narrative structure. The first method, NAR-denoise, mimicked the human rewriting process by introducing noise through deletion and swapping of tokens. It learned to generate high-quality text from noisy input and was further fine-tuned on supervised data. The second method, NAR-reorder, directly addressed the reordering task. It used an input encoding scheme that enabled the model to recognize and handle different sentence structures and coreference chains. Through these two stages of training, the models could generate text in a target narrative order while preserving the story's plot. These methods not only enhanced model performance in terms of fluency and logical coherence but also demonstrated the potential of fine-tuning foundation models to create more interesting and diverse narrative structures.} 

\vspace{0.1cm}
\textbf{Reflection.}
\rv{Logic is enhanced by foundation models, which identify and organize relationships between data facts to create coherent and contextually accurate narratives.
Structure is shaped by foundation models, which enable both interactive creation~\cite{chung2022talebrush} and automatic adjustment of narrative forms~\cite{gangal2022nareor}, allowing for diverse and well-organized storytelling. In narrative visualization, maintaining logical coherence is crucial, as it often involves various forms of reasoning, such as elaboration, similarity, generalization, contrast, temporal sequencing, and cause-effect. These logical structures help create coherent and compelling narratives~\cite{10.1162/0891201054223977,shi2020calliope}. 
 However, the current tools and methodologies do not yet fully leverage foundation models to explicitly clarify and categorize the diverse types of narrative logic.
 
 To address these challenges, future research should \uline{focus on developing methods to explicitly clarify and categorize logical relationships between data facts.}  For example, develop algorithms within foundation models to identify temporal sequences by recognizing time-based patterns in data, or they could detect cause-effect relationships by analyzing correlations and dependencies between variables. Furthermore, future research could \uline{create interactive visual tools that allow users to map the logic.} These tools could work in tandem with foundation models to automatically generate narratives based on the user-defined logic.
}

\vspace{0.3cm}
\subsection{Description}
\begin{wrapfigure}[3]{l}{0.045\textwidth}
    \vspace{-0.4cm}
    \includegraphics[width=0.06\textwidth]{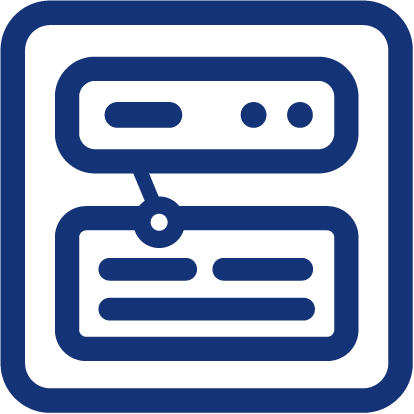}
\end{wrapfigure}

After establishing the logical framework of a narrative visualization, the crafting of detailed descriptions becomes an essential step in adding content to the narration. \mr{Description adds textual details to narrative visualization. It weaves the data insights into a coherent and engaging story to captivate and guide the audience.} Foundation models, with their sophisticated ability to recognize and generate coherent and contextually relevant text~\cite{Patil_Dhotre_Gawande_Mate_Shelke_Bhoye_2023}, play a critical role in this phase. Recent research has
focused on four main aspects: generating titles, captions, annotations, and scripts. 
\vspace{0.1cm}

\textbf{Title} is a line of text or a phrase that typically appears at the top of content, summarizing the main content or theme. The primary purpose of a title is to capture the reader's attention and provide an initial impression of the subject matter. It is often a brief and concise overview of the content\cite{10168257}, which is an important component of effective narrative visualization\cite{10.1145/3290605.3300576}. Foundation models are particularly adept at extracting essential information from visualizations and condensing it into impactful and succinct titles. For example, AutoTitle\cite{10168257} proposed an interactive system designed to generate titles for visualizations by extracting data, computing relevant facts, and using a deep learning model to create natural language titles. The system begins by reverse-engineering the given visualization to extract the underlying data and create atomic facts, which form the foundation for higher-level facts through operations like aggregation, comparison, trend analysis, and merging. A diverse dataset of fact-title pairs is constructed and used to fine-tune the T5 model, generating fluent and diverse titles. 
Users can interact with an interface to upload visualizations, explore generated titles using Radar and RadViz views, and select the most appropriate ones. The interface also includes a representative title view that displays high-quality titles based on the calculated metrics. A user study confirmed the system's ability to produce high-quality titles and validated the usefulness of the metrics.

\vspace{0.1cm}
\textbf{Caption} is typically a brief piece of text placed under images, charts, or graphics to describe or explain these visual elements. The main purpose of a caption is to interpret the picture or chart, provide necessary background information, or explain the relationship between the image content and the main text\cite{liew2022using}. The task of creating a caption for charts using foundation models is a complex one, involving the disciplines of computer vision and natural language processing~\cite{BAI2018291}. \rv{For example, Kantharaj et al.~\cite{kantharaj2022chart} conducted a study collecting 44,096 pairs of chart titles to compare the performance of foundation models in chart captioning. Similarly, Vistext\cite{tang2023vistext} proposed a dataset of 12,441 pairs of charts and captions, which describe the charts' construction, report key statistics, and identify perceptual and cognitive phenomena. This dataset was then used to fine-tune an LLM to generate coherent and semantically rich captions. 

However, this end-to-end generation system limits author involvement in the title generation process, potentially overlooking the author's intended message for the titles. To address this, Intentable\cite{9973198} introduced a mixed-initiative caption authoring system. Intentable allowed users to express their intents through an interactive interface. These intents were encoded as JSON-based `caption recipes'. Utilizing Transformer architectures fine-tuned for this purpose, the system generated accurate and fluent natural language captions based on these recipes. 
Evaluation results demonstrated that Intentable produces more accurate and flexible captions compared to traditional fully automated methods.} 
\rv{Similarly, Liew et al.\cite{liew2022using} used GPT-3 to generate engaging captions for visualizations. The datasets were sourced from Kaggle to create scatter plots, followed by clustering analysis to identify data structures. Effective prompt engineering was employed, involving basic templates, instructional guidance, and an interactive question-and-answer format with the user. The study demonstrated that the incorporation of user interaction and detailed prompt design improved the quality and engagement of GPT-3-generated captions. 

Both Intentable and Liew's research utilize LLMs to generate captions for visualizations but employ different methods. The Intentable system offers highly customized caption generation through user interaction and various intent types, providing flexibility but requiring complex design. In contrast, the system using GPT-3 focuses on rapid caption generation with a tiered approach to prompt engineering, making it quick and user-friendly. The former is ideal for scenarios requiring personalized expression, while the latter suits quick analysis and display needs.} 

\vspace{0.1cm}
\textbf{Annotation} is an additional explanation or commentary provided for text, images, videos, or other narrative content. It can offer background information, explanations, critiques, or a brief analysis of the content\cite{8031599}.  Foundation models can generate annotations, enhancing the information conveyed in narrative visualization by providing more details and context for the audience\cite{bromley2024dash, 10720675}. For instance, Bromley et al. \cite{bromley2023difference} introduced a novel approach that combines feature-word distributions with the visual features and data domain of charts to annotate visual chart features. This feature-word-topic model can identify words associated with vocabulary that are semantically similar but subtly different. Compared to classic automated annotation pipelines like Contextifier\cite{10.1145/2470654.2481374}, foundation models can simplify text mining and image processing techniques. They can even generate annotations through simple prompt engineering. This contrast highlights the advancements foundation models bring to the field, streamlining the annotation process and offering more sophisticated and context-aware outputs with minimal manual intervention.

\vspace{0.1cm}
\textbf{Script} in the context of data storytelling typically refers to a written guide used to organize and narrate the story of data~\cite{segel2010narrative}. Foundation models can generate personalized narrative texts based on specific needs or goals\cite{kalyan2023survey,10.1145/3544549.3583931,keymanesh2022makes,sultanum2023datatales,ye2024storyexplorer}, which means that stories that adapt to different styles, accents, or themes can be generated according to the requirements of the user. This is one of the strongest abilities of the foundation model. For example, Islam et al.\cite{islam2024datanarrative} proposed DataNarrative, a system for automated data-driven storytelling that combines visualizations and text. They presented a multi-agent framework that employs two LLMs: one for understanding and describing the data, generating an outline, and narrating the story, and another for verifying the output at each step. Similarly, Pinargote et al. \cite{pinargote2024automating} proposed a method to automatically generate data narratives in learning analytics dashboards using GPT-3.5. It analyzed and automatically coded classroom meeting transcripts with GPT-3.5 to generate coherent narratives. Furthermore, Ghosh et al. ~\cite{ghosh2023spatio} proposed a new method for analyzing trajectory data called spatio-temporal narratives, interpreting trajectory data as stories. 
Now, foundation models can autonomously generate scripts and stories, meeting a diverse range of user needs and preferences \cite{chung2022talebrush}. 

\rv{\textbf{Reflection.} 
 Title and caption generation by foundation models involve extracting key information from visualizations to create concise and impactful summaries. Annotation is enhanced by foundation models, which generate additional context and explanations to enrich the content. Script generation by foundation models allows for the creation of personalized narrative texts that adapt to different styles and themes.
Due to the potential for hallucinations, the use of foundation models can lead to ethical issues, particularly when the generated narratives are disseminated to a broad audience. These biases can skew the information, potentially reinforcing stereotypes or spreading misinformation, which undermines the integrity and fairness of the content\cite{lingo2023role}. 

To address these challenges, it is essential to adopt a more careful and conscientious approach to using foundation models for narrative generation. Future research could develop advanced algorithms that can detect and correct biases in real-time as the narrative is generated. These algorithms could analyze the content for signs of bias or cultural insensitivity and make adjustments before the narrative is finalized.}

\section{Visualization}
In this section, we review 22 papers related to the narration stage, which can be broadly classified into two categories based on the role of foundation models: (1) visualization generation and (2) visualization authoring. \rr{Specifically}, \rr{generation} pertains to the automated creation of content using foundation models, whereas \rr{authoring} highlights the user's capacity to create and edit narrative visualization content that has already been generated.

\subsection{Generation}
\begin{wrapfigure}[3]{l}{0.045\textwidth}
    \vspace{-0.4cm}
    \includegraphics[width=0.06\textwidth]{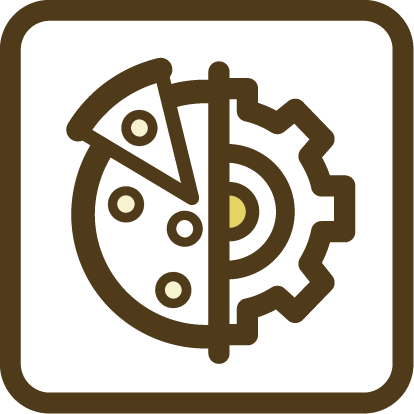}
\end{wrapfigure}
Generation refers to converting data into graphical representations guided by the narrative context and insights derived from the data.
\rv{Foundation} models suitable visualizations to accommodate user-generated visual types and personalization, which is key in narrative visualizations to fit specific user preferences \cite{dibia2023lida}.
Therefore, recent research on visualization generation has focused on three main aspects with foundation models, i.e., charts, images, and animations.

\textbf{Chart}
\rv{refers to a graphical representation of data. Charts use visual elements such as points, lines, and bars to display relationships or trends in the data. Foundation models offer two primary advantages in narrative visualization. On the one hand, }\rv{foundation models can recommend the most suitable chart type based on data insights and narrative purposes\cite{wang2023llm4vis}. 
For example, LLM4Vis\cite{wang2023llm4vis} introduced an interpretable visualization recommendation method. This approach guided ChatGPT to generate comprehensive text descriptions for each tabular dataset, convert these features into vectors, and then employ clustering algorithms for visual recommendations. Consequently, this method eliminated the inefficiencies associated with manual chart selection.}
On the other hand, \rv{foundation models demonstrate the capability to directly generate charts by producing code, a competency validated by multiple studies. For example, Pere-Pau Vázquez et al.~\cite{vazquez2024llms} evaluated the performance of ChatGPT-3 and ChatGPT-4 in generating different types of charts using various visualization libraries such as matplotlib, Plotly, and Altair. Their findings revealed that ChatGPT-4 excelled in producing up to 80\% of the expected charts.

Recent studies have also explored the integration of foundation models' generative capabilities into systems designed to assist users in creating narrative visualizations for practical applications~\cite{10017653,maddigan2023chat2vis,maddigan2023chat2vis2, city31527,zhang2023data, beasley2024pipe}.}
For example, Chat2VIS \cite{maddigan2023chat2vis,maddigan2023chat2vis2} converted natural language into code for appropriate visualization. \rv{The user started by selecting a dataset and entering a natural language query to specify the desired visualization. The system then created prompts that included a data description and a Python code template to assist the LLM in understanding the context and generating precise code. 
In contrast to Chat2VIS, another notable work, LIDA~\cite{dibia2023lida}, has introduced a system that automatically employs a multi-stage, modular approach to generate data visualizations and infographics using foundation models. LIDA used these models in various modules to perform tasks such as data summarization, goal generation, visualization code creation, and the generation of stylized visualizations. 
Compared to Chat2VIS, LIDA is a more comprehensive end-to-end visualization generation system. LIDA allows user intervention at every stage. It also features robust multilingual interfaces for optimizing, explaining, and evaluating charts, making it suitable for novice users and those with technical backgrounds.
The previously mentioned research uses online models, which might have suffered from inherent hallucination issues associated with foundation models. In contrast, ChartGPT~\cite{tian2023chartgpt} developed a visualization-specific LLM aimed at improving chart recommendations. The approach entailed the development of a comprehensive dataset comprising abstract statements paired with their corresponding charts. Although foundation models can assist in generating charts, current research predominantly focuses on the creation of individual charts, whereas narrative visualizations often require multiple logically connected charts.
} 

\textbf{Image} refers to the illustrations and graphics used to enhance, interpret, or supplement data narratives. \bl{These images may include characters, scenes, emotional expressions, or other visual elements that support the storytelling aspect of the narrative.} Foundation models like DALL-E\cite{ramesh2022hierarchical,ramesh2021zero} have revolutionized this domain by enabling the automatic creation of images from textual descriptions. Such models enrich narrative content by producing images that align closely with the underlying story, thereby elevating its appeal and comprehensibility\cite{heidrich2024generative, 10.1145/3613905.3651111}. \rv{Schetinger et al.~\cite{di2023doom} highlighted that generative text-to-image models present substantial opportunities in visualization processes. This included the design of visualizations, the creation of mood boards, and the generation of explanatory images. For example, Ghosh et al.\cite{ghosh2023spatio} leveraged  DALL-E to provide visual explanations for generated semantic trajectory visualizations, offering a more intuitive way to understand and interpret mobility patterns. This method provides a richer and more comprehensive depiction of the scenarios under study. 
Foundation models can generate images through text-to-image or image-to-image methods, effectively enriching the visual aspect of narrative visualizations. Future research could explore how to generate images in different styles based on user preferences, further enhancing user engagement and appeal. }
\begin{figure*}[htbp]
    \centering
    \includegraphics[width=1\textwidth]{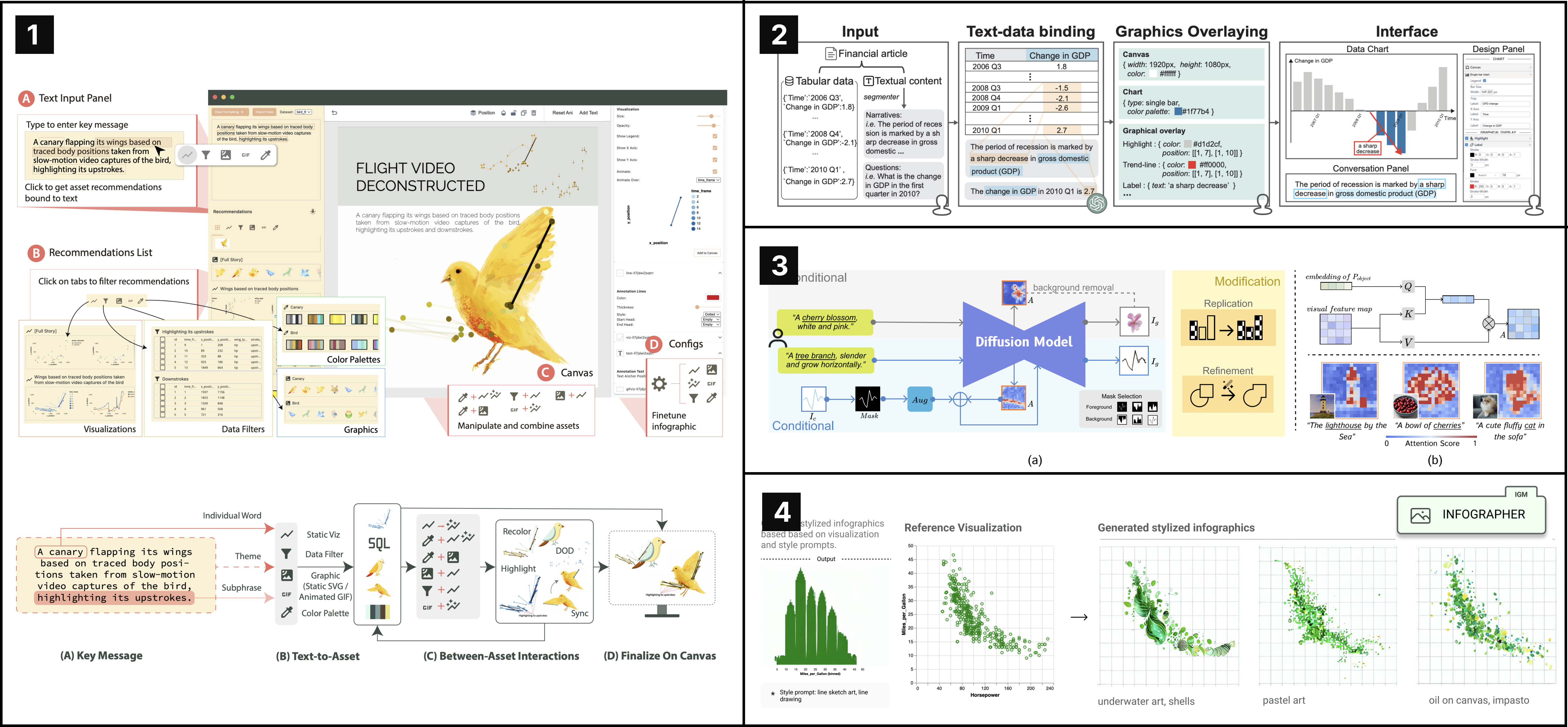}
    \caption{ Selected examples of editing. \mr{(1)~Epigraphics: message-driven infographics authoring~\cite{10.1145/3613904.3642172}. (2)~FinFlier: layer visual elements onto charts of financial narrative visualizations~\cite{10787087}.(3)~ChartSpark: embedding semantic context into charts with text-to-image generative model~\cite{xiao2023let}.  (4)~LIDA: generate stylized graphics based on visualizations for infographics~\cite{dibia2023lida}.   }}
    \label{fig:authoring}
\end{figure*}
\textbf{Animation} is a technique that brings static images and graphics to life visuals by creating the illusion of movement\cite{beckerman2003animation}.
\rv{In the context of narrative visualization, data video is an important form of narrative visualization, integrates visualization with animation to convey data-driven stories\cite{chen2023does}. Traditionally, creating a data video required professional skills. However, foundation models offer two primary advantages in narrative visualization.

The first advantage is that foundation models can connect narrative text with visual elements.  For example, Ying et al.\cite{ying2023reviving} used Graph Neural Networks to parse static SVG charts and extract their data and visual encodings. They then utilized GPT-3 to generate descriptive and insightful audio narrations based on the charts. By adding appropriate animations to the chart elements and synchronizing them with the narrations, static charts were transformed into dynamic charts with enhanced interactivity and information delivery.
Similarly, Data Player\cite{shen2023data} proposed a method for automatically generating data videos. This technical solution first extracted data by parsing static SVG charts and mapping visual elements to rows in a data table. Then, it used GPT-3.5-turbo to semantically match narrative text segments with data table rows, creating links between text and visual elements.
Both studies involved extracting data from static charts and transforming it into a more manageable format. They utilized foundation models for natural language processing to generate descriptive narrations and establish links between text and visual elements, and employed programming or constraint-solving solutions to generate animation sequences synchronized with narrations. 

The second advantage is that foundation models can create animations by generating interpolations between the start and end images. Generative Disco\cite{liu2023generative} used foundation models to integrate visual elements with music rhythm to generate animations. It utilized models such as GPT-4 to generate prompts that encompassed text and visual objectives, aligning these with the lyrics and rhythm of the music to ensure coherence with the musical content. For the animation interpolation process, Generative Disco employed Stable Diffusion to create transitions between the start and end images, analyzing music beats and dynamics to produce animation effects synchronized with the musical rhythm. In general, current animation generation techniques focus primarily on connecting individual images and charts to animations using interpolation or rule-based syntax. However, there is a lack of examples in which foundation models are used directly to generate animations.}

\rv{\textbf{Reflection.} Charts can be generated by foundation models through code production, creating visualizations that represent data. Images can be produced by leveraging the text-to-image capabilities of foundation models, enriching the content of data comics and other narrative visualizations. Animations are created when these charts and images are connected according to a narrative structure, with transitions such as interpolation applied, effectively conveying complex data stories.
Although foundation models perform well in generating most standard charts, they still face notable difficulties when producing more complex visualizations, such as Sankey diagrams and composite charts. Additionally, another challenge arises when foundation models are required to configure multiple visual variables simultaneously~\cite{lin2024observable}. For example, when multiple variables need to be adjusted, the generated code may not execute correctly, resulting in errors.

To address these challenges, future research should focus on \uline{improving foundation models in designing more complex and composite charts }\cite{shen2023data}. Overcoming limitations in handling comprehensive visualizations will enhance the utility of foundation models.
In addition, most methods currently rely on linking charts to create animations. However, with the advancement of video models, \uline{video foundation model is expected in the future that is fine-tuned specifically for narrative visualizations.}}

\rv{\subsection{Authoring}}
\begin{wrapfigure}[3]{l}{0.045\textwidth}
     \vspace{-0.4cm}
    \includegraphics[width=0.06\textwidth]{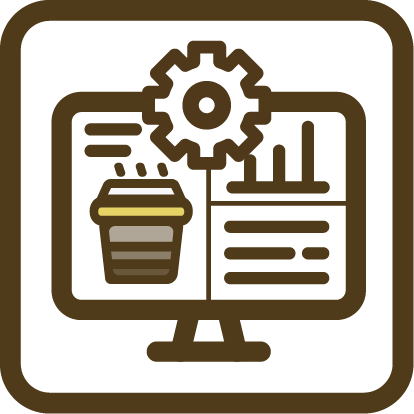}
\end{wrapfigure}
\noindent 
\mr{Authoring is the process of transforming individual materials—such as narratives, charts, and images—into a cohesive and compelling narrative visualization~\cite{10.1145/3613904.3642172}. It involves editing and enhancing each element to align with user needs while establishing connections among them to create a unified flow.}
\mr{In this process, foundation models primarily assist users in establishing relationships between materials and correcting errors. Additionally, their generative capabilities enable the editing of materials to enhance clarity and aesthetics and the effectiveness of information communication.}
Therefore, recent research in this domain has focused primarily on three main aspects: encoding, embellishment, and layout.

\rv{\textbf{Encoding} refers to the specification used to map data attributes to visual elements. It enables data to be represented visually, allowing people to understand and analyze the data intuitively~\cite{guo2009flow}. 
Previous research has aimed to help users detect and correct encoding errors in visualizations. For example, Chen et al. proposed VizLinter~\cite{9552878}, a technical solution that incorporates a visualization linter to detect errors in visualizations and a visualization fixer that employs a linear optimization algorithm to automatically suggest and apply corrections. More recently, Lo et al.\cite{lo2024good} evaluated the abilities of foundation models to identify improper encoding. In this study, multimodal foundation models detected chart errors, particularly encoding errors, by employing prompts of varying complexity, ranging from simple direct questions to complex chain-of-thought strategies. The researchers utilized multi-round conversation prompts to guide the models through a progressive analysis of chart issues. Experimental results demonstrated that foundation models performed significantly better in identifying encoding errors when guided by chain-of-thought prompts. Therefore, this study highlighted the potential of multimodal foundation models to enhance encoding accuracy and reliability through advanced prompt engineering techniques. By employing chain-of-thought reasoning, foundation models can deduce encoding errors. Future research could explore whether these models can be further guided through dialogue to generate code that corrects encoding errors.}

\rv{\textbf{Embelishment} refers to the addition of decorative and ornamental elements in the process of data visualization to enhance visual impact and user comprehension\cite{9903511}.} Foundation models can be leveraged to analyze data and context, enabling the recommendation of suitable graphical elements such as icons and illustrations to enhance the visual impact and information delivery~\cite{10787087}. 
\mr{For example, Zhou et al.\cite{10.1145/3613904.3642172} proposed Epigraphics, an authoring tool that enhances the aesthetic cohesiveness of the materials, as shown in Fig. \ref{fig:authoring} (1). The system recommends visualizations, graphics, data filters, color palettes, and animations based on text inputs, supporting interactions like recoloring and highlighting. GPT-3.5 analyzes user input to generate chart specifications and SQL for data filtering. Sentence-BERT optimizes recommendations by calculating text-graphic similarity, while BLIP provides image captions. Adobe Firefly creates images and suggests colors, collectively improving infographic content and presentation.
In another study, FinFilter utilized GPT-3.5 to automate the graphical overlays for financial narrative visualizations, as shown in Fig. \ref{fig:authoring} (2). The study summarized common graphical overlays and narrative structures from a survey of 1,752 layered charts. By applying prompt engineering techniques, financial knowledge was integrated into GPT-3.5, enabling it to efficiently identify complex structures and link terminology with data. Foundation models were  used to automatically generate layered charts, optimizing overlay positions and colors to enhance visualization effectiveness.

Recently, research also focused on stylized and pictorial visualizations based on text-to-image or image-to-image models. These approaches replace the foreground and background of charts with semantically relevant graphics. For example, Viz2viz\cite{wu2023viz2viz} employed a combination of generative diffusion models and image processing techniques to transform traditional visualizations such as bar charts, area charts, pie charts, and network graphs into highly stylized forms. 
ChartSpark \cite{xiao2023let} introduced a three-stage framework—feature extraction, generation, and evaluation—to embed semantic context into charts using text-to-image generative models, as shown in Fig. \ref{fig:authoring} (3). MPNet and Word2Vec facilitated keyword extraction and thematic context. The Frozen Latent Diffusion Model integrated data and semantic context. An evaluation stage ensured chart fidelity through pixel-wise comparison and edge detection, enhancing interpretability and credibility.
In another study, LIDA \cite{dibia2023lida} generated stylized graphics to refine the infographics by using a user-editable library of visual styles and the text-conditioned image-to-image transformation with diffusion models, as shown in Fig. \ref{fig:authoring} (4). }

\rv{\textbf{Layout} is the underlying semantic structure that links graphical elements to convey the information and story to the user\cite{10.1145/3313831.3376263}.  Recent advancements in foundation models have enabled the parsing of natural language instructions to facilitate chart layout adjustments. ChartReformer\cite{yan2024chartreformer} achieves this adjustment by parsing user natural language instructions, using a pre-trained Visual Language Model(VLM) to extract the current visual properties and layout information of the chart, generating new drawing parameters, and invoking charting libraries to redraw the chart. The specific steps include: first, the system parses user instructions using NLP techniques to understand the desired chart layout adjustments. Next, the system uses the VLM to analyze the input chart image and extract its current attributes, such as axes, gridlines, and legend positions. Based on the parsed user instructions, the system is able to generate new drawing parameters. Finally, the system invokes charting libraries like Matplotlib or Vega-Lite to apply the new drawing parameters and redraw the chart such that the layout is adjusted according to the user's instructions. This process combines natural language processing, visual language models, and advanced charting libraries to achieve precise and user-driven chart layout editing. Similarly, the DynaVis system\cite{10.1145/3613904.3642639}, powered by the GPT-3.5, enabled users to edit various aspects of visualizations using natural language commands, including chart layout, visual properties, data filtering and transformation, and axis range adjustments. Current layout editing tools primarily focus on basic adjustments such as modifying axes and legends. However, the overall distribution of visual elements, the sequence in which charts are presented, and the spatial organization within a visualization can greatly influence how users interpret and engage with the data. Future research could delve into more sophisticated layout strategies that optimize the user experience.}

\rv{\textbf{Reflection.}
\bl{Foundation models enhance the authoring process by improving encoding, embellishment, and layout in visualizations. For encoding, these models help detect and correct errors, ensuring accurate data-to-visual mappings. For embellishment, they recommend and generate graphical elements like icons and illustrations, enriching the visuals through advanced data analysis and contextual understanding. Finally, for layout, foundation models enable precise adjustments by parsing natural language instructions, extracting visual properties, and applying new parameters via charting libraries, allowing for user-driven editing.}

\bl{However, current authoring tools exhibit several limitations. First, the range of supported chart types and narrative visualization formats is limited. \uline{Future authoring tools should expand to include a broader variety of chart types and narrative forms to better represent complex data and tell diverse stories.} Second, the customization options in current tools are inadequate, restricting users’ creative expression. To address this problem, \uline{incorporating sketching or hand-drawn authoring capabilities could allow users to adjust visual elements more freely}~\cite{10.1145/3613904.3642172}. Finally, layout design in current tools lacks effective control over the overall information flow and the ability to guide the viewer's reading sequence. \uline{Future research could adopt more innovative layout styles, such as spiral or star arrangements}, to improve narrative readability and more effectively guide the viewer’s attention ~\cite{10.1145/3313831.3376263}.}
}


\vspace{8mm}
\section{Interaction}
We reviewed 15 papers related to the interaction stage, which can be broadly classified into two major categories based on the role of foundation models: (1) chart interpretation and (2) data navigation. This section will discuss both categories in detail.

\begin{figure*}[htbp]
    \centering\includegraphics[width=1\textwidth]{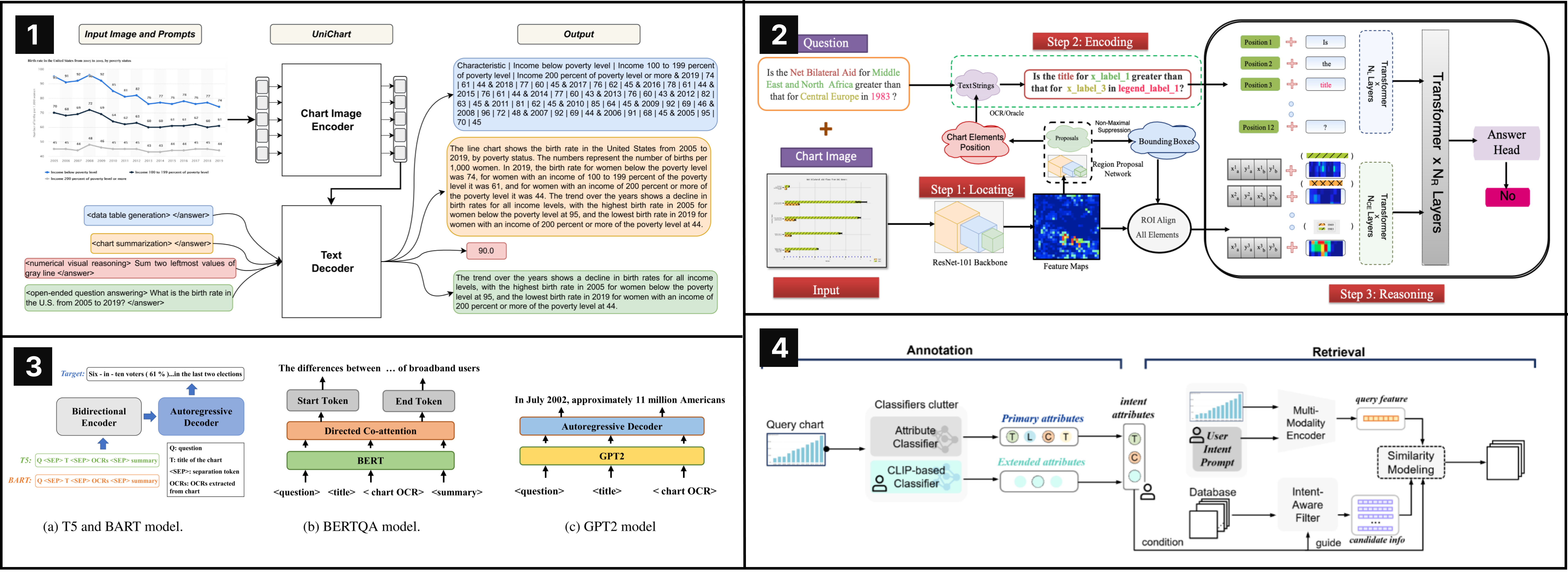}
    \caption{ Selected examples of Chart Understanding. Chart Understanding is divided into three tasks: \textbf{Chart Summarization}: (1) Unicharts: a model that processes visual elements through a chart encoder and text decoder to generate natural language summaries\cite{masry2023unichart}; \textbf{Chart Question Answering}: (2) STL-CQA: a structured Transformer model to localize chart elements and perform cross-modal reasoning~\cite{singh-shekhar-2020-stl}, (3) OpenCQA: Multiple foundation models were fine-tuned to answer open-ended questions with charts\cite{kantharaj-etal-2022-opencqa}; \textbf{Chart Retrieval}: (4) WYTIWYR: a user intent-aware framework with multimodal inputs for visualization retrieval\cite{xiao2023wytiwyr}.
 }
    \vspace{-2mm}
    \label{fig:comprehension}
\end{figure*}
\subsection{Chart Interpretation}
\begin{wrapfigure}[3]{l}{0.045\textwidth}
    \vspace{-0.4cm}
    \includegraphics[width=0.06\textwidth]{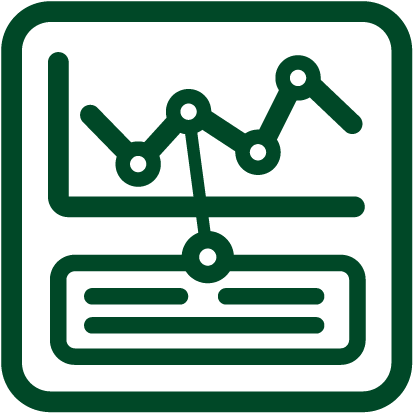}
\end{wrapfigure}
Chart interpretation involves the automated ``reading'' of narrative visualizations in a way that simulates human perception~\cite{wang2021survey}. Understanding important patterns and trends from charts and answering complex questions related to charts can be cognitively demanding. Furthermore, accessibility challenges exist for people with visual impairments in understanding charts. To address these challenges, recent research has focused on two main aspects: chart derendering and chart understanding~\cite{masry2023unichart} based on foundation models.

\textbf{Chart Derendering} 
\rv{is the process of converting charts into tables~\cite{cheng2023chartreader,meng2024chartassisstant}.} This process presents significant challenges due to the variety of chart types and the complexity of their components. 
Existing approaches are heavily based on heuristic rules \rv{for detecting chart components}, which require substantial domain knowledge to develop~\cite{Luo_2021_WACV}.
\rv{Foundation} models, however, have shown remarkable proficiency in plot-to-table tasks without explicitly recognizing chart structures and components~\cite{cheng2023chartreader}, which led to their increase in automatic content recognition and extraction.  
\rv{A notable example was MATCHA\cite{liu2023matcha}, which was designed based on the encoder-decoder structure of the Pix2Struct model and achieved chart-to-table conversion through end-to-end training. The model employed a Transformer architecture with 300 million parameters and was fine-tuned on real-world chart-table pairs to generate linearized tables. 
}
\rv{On the basis, Deplot\cite{liu2022deplot} further fine-tuned MATCHA by focusing exclusively on the task of chart-to-table conversion. To better accommodate complex charts with numerous data, Deplot utilized a longer sequence length, enhancing its performance in chart derendering tasks. By decomposing the charts into tables, users could input the resulting tables along with the questions as text prompts, which can then be further processed by the foundation models for enhanced chart understanding.}

\rv{\textbf{Chart Understanding}} aims to attain a comprehensive interpretation of visualizations and tackle complex analytical tasks.\cite{wu2021ai4vis}. Numerous studies have advanced this field by collecting extensive datasets, fine-tuning foundation models, and integrating multiple chart understanding tasks to improve the performance of narrative visualizations\cite{han2023chartllama,zhou2023enhanced,10670418,10766490}. Foundation models can solve three downstream tasks: (1) Chart Summarization\cite{rahman2023chartsumm,kantharaj2022chart}, (2) Chart Question Answering~\cite{masry2022chartqa} \rv{and (3) Visualization Retrieval\cite{xiao2023wytiwyr}}. 

Chart Summarization aims to articulate the important data points and trends presented in a chart using natural language\cite{rahman2023chartsumm}. 
This task could be beneficial for readers seeking to quickly grasp the main points of a chart \cite{kantharaj2022chart}. Leveraging their reasoning and generative abilities, \rv{foundation} models can produce concise and accurate textual summaries that capture the essential information and trends in charts. For example, \rv{Unichart\cite{masry2023unichart} comprised a chart encoder to process the relevant charts, text, data, and visual elements of the chart, and a text decoder to generate the corresponding natural language output, as illustrated in Fig. \ref{fig:comprehension} (1). To enhance chart comprehension and reasoning capabilities, Unichart was trained on a large corpus of 611K charts with various pre-training tasks, including data table generation, data value estimation, numerical and visual reasoning, open-ended QA and chart summarization. Although Unichart excelled in generating concise and accurate textual summaries, it did not perform as well as specialized QA systems in handling open-ended questions or complex logical reasoning. }

Chart Question Answering is a task to take a chart and a question in natural language as input to generate the desired answer as output\cite{https://doi.org/10.1111/cgf.14573,10555321,gao2024fine}. \rv{Foundation} models are capable of understanding and processing user queries to provide relevant and accurate answers. These models go beyond surface-level information analysis, employing logical reasoning and contextual analysis to provide deeper insights in response to user inquiries. For example, STL-CQA\cite{singh-shekhar-2020-stl} employed a structured Transformer model to localize and encode chart elements for answering natural language questions about charts. STL-CQA's technical approach involved three key steps: first, detecting and localizing chart elements using a Mask R-CNN model; second, dynamically encoding questions by replacing specific words with standardized tokens; and finally, using a structured Transformer module for chart structure understanding, question understanding, and cross-modal reasoning, as shown in Fig. \ref{fig:comprehension} (2). 
However, STL-CQA faced limitations when dealing with open-ended questions or tasks requiring larger parameter models.\rv{ In contrast, OpenCQA~\cite{kantharaj-etal-2022-opencqa} leveraged larger parameter models to handle open-ended questions about charts, enabling deeper logical reasoning and contextual analysis. OpenCQA's technical approach encompassed three task settings: using charts and their accompanying full articles, providing only relevant paragraphs, and generating answers based solely on the charts. As shown in Fig. \ref{fig:comprehension} (3), OpenCQA's dataset included 7,724 open-ended questions and descriptive answers, and multiple foundation models were fine-tuned. Through various model architectures and rigorous evaluation methods, OpenCQA showed the potential to generate detailed descriptive responses. However, it still faced challenges in handling complex logic and mathematical reasoning. Although this method required high computational resources, it offered deeper insights when addressing complex problems.} 

\rv{Chart Retrieval refers to the process of finding charts from a set of charts that meet specific query conditions or requirements. Chart retrieval can efficiently locate relevant visualizations from vast datasets, enabling users to quickly access and utilize the most pertinent charts for their needs. In this task, foundation models can be used to map textual and visual features into a shared embedding space, which enables zero-shot classification of the query charts. For example, WYTIWYR\cite{xiao2023wytiwyr} introduced a user intent-aware framework for chart retrieval through multi-modal inputs. This framework involved two main stages: annotation and retrieval. In the annotation stage, a deep neural network classified query charts and decoupled their visual attributes using the CLIP model for zero-shot classification. In the retrieval stage, an intent-aware filter sifted through the chart database based on user-selected intent attributes, while multi-modal encoders generated joint feature representations from the inputs, as shown in Fig. \ref{fig:comprehension} (4). WYTIWYR integrated user intent into the chart retrieval process, providing highly customized results. However, this approach required substantial computational resources, and designing effective text prompts remained a challenge.}

\rv{\textbf{Reflection.} Chart Derendering leverages foundation models to convert charts into tables through advanced neural networks, effectively bypassing the need for explicit recognition of chart structures and components. Chart Understanding utilizes foundation models to process visual elements and textual data, enabling tasks such as summarization, question answering, and retrieval, which allow for comprehensive interpretation of narrative visualizations.
\bl{However, current techniques are primarily focused on derendering and understanding individual elements, such as single charts, which are often applied in general visualization. In contrast, narrative visualization requires a cohesive flow of context and relationships between multiple charts, as these connections are essential for building a comprehensive and meaningful story. }

To address these limitations, a promising area for future research lies in \uline{exploring how to develop a holistic approach that cohesively integrates multiple charts.} This involves not just interpreting individual charts, but also understanding how they interact with each other to form a unified story. Foundation models can be further developed to analyze and interpret these interrelationships. Users are able to obtain a more comprehensive and logically coherent understanding of the entire narrative visualization~\cite{https://doi.org/10.1111/cgf.14573}. For example, future research can explore the design of contextual links between charts, where hovering over or clicking on an element in one chart provides insights or highlights relevant data in other charts. This would enhance the user's ability to see connections across different visual elements and deepen their overall understanding of the narrative.
}
\subsection{Data Navigation}
\begin{wrapfigure}[3]{l}{0.045\textwidth}
    \vspace{-0.4cm}
    \includegraphics[width=0.06\textwidth]{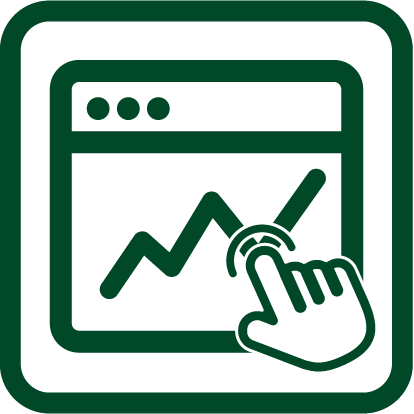}
\end{wrapfigure} In narrative visualization, data navigation is the process where users browse and explore data via a visual interface. This includes activities like selecting different data views, zooming, and scrolling to access various parts of the data, and altering the data representation using interactive elements like sliders, buttons, or drop-down menus\cite{chen2023beyond}.

\textbf{Navigation} \rv{refers to how individuals engage with visualizations to explore data\cite{wang2021survey}.} \rv{Foundation models have enhanced user interaction with narrative visualizations by enabling more intuitive navigation and exploration through natural language queries. These models allow users to interact with visualizations simply by adjusting the prompts, which reduces the barriers to accessing and interpreting complex data.
Furthermore, foundation models provide the ability to customize data views based on individual user interests and requirements. They adapt data displays according to user-specified parameters and can even generate code to offer a more personalized experience. For example, Chen et al.~\cite{chen2023beyond} demonstrated how GPT-4 interacted with charts by sending JavaScript events, generating code to update axes, resetting zoom levels, and mimicking hover events for tooltips. This extended the adaptability to tailoring data insights for various audiences, such as GPT-4, which can generate audience-specific reports for different entities such as the governments of African countries, the United Nations, and primary school students.

In addition to natural language interaction, graphical methods offered new ways to express analytical intent. InkSight\cite{10296056} explored a novel interaction method that allowed users to directly draw paths on data visualization charts to express their analytical focus. This intuitive method allowed users to select subsets of data or patterns of interest with more precision, which were then automatically annotated by the system. InkSight's approach revolved around capturing user intent and automatically generating documentation that reflected this focus. By sketching directly on the charts, users expressed their analytical intent, which the system interpreted and converted into natural language documentation using GPT-3.5. This documentation could be further refined through editing, making it precise, and easy to understand, and enhancing the efficiency of recording and sharing data analysis results.}

\rv{\textbf{Reflection.} Navigation is enhanced by the foundation models through the use of NLP and adaptive algorithms, allowing users to intuitively interact with and explore data by adjusting prompts and customizing data views. Although foundation models have shown the potential to lower barriers to audience engagement and interaction, previous research indicates that it remains challenging to describe interactions. Non-experts, in particular, often find it difficult to naturally express complex interactive functions using natural language\cite{vazquez2024llms}. Firstly, non-expert users are generally unfamiliar with the technical language required to effectively guide the model. Foundation models, while powerful, still require precise and clear prompts to generate the desired visualizations. However, users without a background in narrative visualization may struggle to articulate their intentions in a way that the model can accurately interpret. This gap between the user's natural language and the model's interpretive capabilities often leads to misunderstandings, resulting in interactions that do not fully meet the user's needs.
Secondly, when presented with a narrative visualization, non-expert users often find themselves unsure of how to proceed. The complexity of interacting with and manipulating the visualization can be overwhelming, especially when the visualization includes multiple layers of data or advanced interactive elements.

To address these limitations, future research should focus on \uline{providing users with more explicit guidance on operational methods.} Research should explore ways for users to learn these methods. Potential learning pathways include \cite{segel2010narrative}:
\textbf{\textit{Explicit Instruction:}} Offering clear, step-by-step operational explanations to help users quickly grasp basic operations.
\textbf{\textit{Tacit Tutorial:}} Implementing implicit tutorials and progressive guidance to allow users to naturally master complex features during usage.
\textbf{\textit{Initial Configuration:}} Providing a well-designed initial setup that is easy to understand and operate, reducing the difficulty of getting started.
By combining the generative capabilities of foundation models with these interaction design strategies, future narrative visualization tools can serve user needs better and facilitate the widespread application of narrative visualization technology.
}

\section{Evaluation}
\mr{In this section, we discuss existing methods for evaluating the reliability and quality of narrative visualizations created by foundation models. Furthermore, we propose future research directions to offer valuable insights for researchers and practitioners in the field of narrative visualization.}

\subsection{Methods of Evaluation}
\mr{Although foundation models provide convenience in creating narrative visualizations, their widespread application faces some potential challenges. One major issue is the variability in how these models can be instructed, as different prompts can lead to divergent and unpredictable results of the process. Moreover, the presence of hallucinations, where models generate inaccurate or non-sensical content even in tasks as structured as visualization generation, further complicates their use\cite{podo2024vi}. 

Recently, several studies have emerged to evaluate visualizations created by foundation models \cite{chen2023beyond, zhu2024large, lo2024good} through various human evaluations from various perspectives to identify their strengths and weaknesses. However, in practical applications, automated evaluation methods and a comprehensive set of metrics are necessary to evaluate the visualizations generated by these models effectively.
For example, Podo et al. \cite{podo2024vi} proposed a conceptual framework called EvaLLM, which systematically dissects and models the evaluation of data visualizations generated by LLMs. This framework consists of five layers. From bottom to top, they are the \textbf{Code layer} (evaluating the consistency of the visualization within the code environment), the \textbf{Representation layer} (assessing the semantic aspects of the visualization representation), the \textbf{Presentation layer} (evaluating the data presentation quality of the visualization from the perceptual perspective), the \textbf{Informativeness layer} (measuring the intrinsic quality of the visualization), and the \textbf{LLM layer} (evaluating the generation strategy and the significance of the visualization). Each layer includes different levels that support the evaluation process, detailing aspects in terms of evaluation scope, importance, implementation approach, evaluation affordability, semantics, and human-machine ratio. 

To enhance the generalizability of the evaluation, Chen et al. \cite{chen2024viseval} proposed a dataset that includes 2,524 representative queries, covers 146 datasets, and is paired with accurately labeled ground truth. They also introduced a comprehensive automated evaluation method encompassing multiple dimensions:
The\textbf{ validity checker} ensures the code generates visualizations by first running it in a sandbox to avoid crashes, then checking for essential code snippets to render the visuals.
The\textbf{ legality checker} verifies if the visualization matches the query by extracting key details (e.g., chart type, data) and checking their accuracy and order using meta-information.
The\textbf{ readability evaluator} evaluates visualization clarity through layout and scale/tick checks. It assigns a readability score (1–5) based on factors like titles, labels, and colors.

\textbf{Reflection.} These evaluation frameworks effectively decompose the tasks involved in assessing data visualizations created by foundation models. However, beyond traditional evaluation criteria, narrative visualizations demand a more comprehensive framework that incorporates new metrics specifically related to narratives, such as narrative coherence, narrative relevance, and audience engagement. For instance, while it is important to assess the technical accuracy of the visualizations, assessing the narrative logic is equally important. This includes ensuring that the story flows coherently, the insights are effectively communicated, and the data meaningfully supports the narrative. Furthermore, the engagement factor should be considered, as a good narrative visualization must not only present data but also captivate the audience’s attention. Evaluation should consider elements like the clarity of storytelling, as well as the emotional or cognitive impact the visualization has on the viewer.}

\rv{\subsection{Research Opportunities of Evaluation}
Although effectively evaluating and enhancing the narrative quality of these models remains a critical challenge, relatively comprehensive visualization evaluation methods for data visualization are available. However, evaluation methods specifically tailored for narrative visualization require further refinement. Additionally, improving evaluation efficiency and enhancing user trust through evaluation are also future research directions.

\textbf{Using the Foundation Model as an Evaluation Method.}
Future research can explore how to use the generative model itself to evaluate its narrative output\cite{podo2024vi}. Leveraging the self-evaluation capabilities of the model can lead to more precise and efficient assessments of narrative quality. This involves training the model to understand and evaluate the coherence, logic, and linguistic quality of its own generated content, thereby enhancing the overall standard of the narratives produced. This approach can reduce the subjectivity of human evaluation and improve the consistency and reliability of assessments.

\textbf{Visualizing Evaluation and Error Correction.}
Visualizing the evaluation and error correction processes of generated narratives can help users more intuitively understand the quality and issues in the model output\cite{xiao2023let}. Future research could develop more tools that allow users to see evaluation results and perform real-time error correction. This not only aids users in better understanding and utilization of the model but also promotes the improvement and optimization of the model itself. Through visualization, users can more easily identify and correct errors in the generated content, enhancing the quality and acceptability of the narrative output.}

\vspace{5mm}
\section{DISCUSION}
\rv{In the previous sections, we analyzed the application of foundational models to automate narrative visualization using the defined reference model. Having identified future research directions for each task, we now present a comprehensive analysis of prospective research avenues for the overall process.

\mr{\textbf{Crafting Narrative Visualizations with Agent-based Workflow.} Recent advancements in foundation models have facilitated the development of autonomous agent-based systems that simplify the data storytelling workflow. In this framework, foundation models serve as the cognitive core of agents, defining distinct agent roles and allowing them to perceive, decide, collaborate and act. This automation of tasks enhances the efficiency of narrative visualization creation\cite{shen2024data} and positions multi-agent systems for narrative visualization as a promising avenue for future studies.

\uline{Agent Role Definition and Task Allocation.} Foundation model serves as the brain of autonomous agent-based systems, facilitating the assignment of specific roles to manage various aspects of the narrative visualization workflow. Based on specific tasks drawing from our reference model, foundation models could design efficient systems that decompose the narrative process into manageable sub-tasks, and assign appropriate roles to agents specialized in handling particular aspects of these tasks~\cite{shen2024data}. For example, 
in an automated data insight and narrative generation system, foundation models such as LLMs can assist in mimicking the human creative process for storytelling. They can establish a data analyst role to handle data exploration, alongside roles such as scripter and reviewer for story production, \rr{They can establish a data analyst role to handle data exploration, along with roles such as scripter and reviewer for story production, a concept similar to the discussion of the DataNarrative~\cite{islam2024datanarrative} system in the \textbf{Script}.  
Finally, an editor role can be designated for the presentation of the story.}
For a visualization system, we can assign a designer agent that utilizes image-based models or multimodal models to create visual representations, ensuring that the narrative remains compelling and that the visualizations are consistent with the story.


\uline{Information Flow and Collaboration between Agents.} 
Effective communication and collaboration between agents are vital for maintaining a seamless crafting workflow. Foundation models can ensure that each agent converts input data into a format that can be easily processed by the models themselves, thereby strengthening the contextual foundation for subsequent tasks. Additionally, \rr{as we discussed in the \textbf{Logic}}, prompt engineering for foundation models is particularly important, as it is essential to define the input-output formats and the chains of thought for each agent to ensure a comprehensive workflow.

}

\textbf{Affective Narrative Visualization Design.} Substantial research demonstrates that emotionally rich designs can significantly influence narrative visualization\cite{lan2023affective}. Recent studies indicate that foundation models are now capable of analyzing and generating emotional content\cite{zhang2024refashioning}. Therefore, future research should focus further on enhancing the affective computing capabilities within foundation models and exploring novel applications that fully leverage the role of emotions in storytelling. Furthermore, by integrating foundation models with existing design spaces, researchers can create more engaging and emotionally compelling narrative visualization. Based on these insights, future research directions should address the following areas:

\uline{Personalized Emotional Experience.}  Designing systems that can tailor the emotional tone of narratives based on individual user preferences and profiles is a promising direction for enhancing personalized emotional experiences\cite{10.1145/3382507.3418884}. \rr{To craft a personalized visualization, user modeling and adaptive algorithms are necessary. As discussed in the \textbf{Navigation}, foundation models have the ability to customize data views according to the unique interests and needs of users. Through prompt engineering, foundation models adjust the narrative visualization based on user-defined parameters.}
For example, if a user prefers uplifting and motivational stories, the system can prioritize content with positive emotional tones. In contrast, for users who enjoy dramatic and intense narratives, the system can introduce elements that evoke stronger emotional responses. This personalized approach enhances the relevance and emotional impact of the narrative, making it more engaging for each user.

\uline{Emotion-Driven Storytelling.}  Understanding emotions in narratives to make models more human-like is a crucial direction for future research on foundation models. By improving the model's ability to comprehend and convey emotions, the generated narratives can become more vivid and impactful. Particularly, research can focus on collecting emotionally rich datasets to fine-tune foundation models\cite{Yang_2024_CVPR,Yang_2023_ICCV}, or adjusting existing model architectures or prompts to help models accept and analyze emotional input, thus enhancing the naturalness of emotional expression in storytelling\cite{shen2023data}.}

\rv{\textbf{Interaction in Narrative Visualization.}
Interaction plays a crucial role in narrative visualization by bridging the gap between data representations and user experiences. Although tools like Tableau and APIs such as d3.js can create interactive visualizations, these efforts often fail to effectively facilitate narrative communication. Moreover, constructing interactive elements, even for simple charts, can be complex and time-consuming\cite{6634113}. Therefore, leveraging foundation models to simplify and enhance the crafting of interaction in narrative visualization is a key focus for future research. The following directions merit exploration:

\uline{Enhancing User Experience through the Natural Language-based Interaction.}  \rr{In the previous discussion on various tasks involved in creating narrative visualizations, such as \textbf{Data Exploration, Generation}, and \textbf{Chart Understanding}, we emphasize that natural language-based interaction eliminates the need for users to master specialized technical skills. Users can simply pose questions or commands in natural language, and the models can accurately interpret and respond. This simplified interaction reduces the learning curve, making the creation of narrative visualizations more accessible to a broader audience. } The natural language processing capabilities of foundation models enable them to comprehend and generate a wide range of semantic information, allowing them to handle complex user commands and natural language queries. Foundation models can not only understand the user's intent but also generate precise responses and recommendations tailored to specific needs\cite{chen2023beyond}.  

\uline{Enhancing User Interaction through Exploratory Interaction.} \rr{Some tools discussed in the \textbf{Generation} usually pre-determine a template, regardless of user input, which results in a limited variety of outcomes.} While this end-to-end approach offers simplicity and clarity in certain applications, it restricts user interaction and limits the diversity of the narrative. As the application of foundation models in visualization deepens, there is growing interest in how user interaction with these models can lead to diverse narrative paths and multiple visualization forms.

Future research could prioritize the design and implementation of adaptive narrative structures. These structures would lead to different perspectives and outcomes with different user interactions, thereby improving the complexity and richness of narrative visualizations. This approach presents not only a technical challenge, but also a deep understanding of user behavior and cognitive processes to ensure that the generated narratives meet user expectations and effectively convey critical information.}

\vspace{2mm}
\section{conclusion}
In this study, we \rv{conducted a systematic} review of 77 papers to study how \rv{foundation} models progressively engage in narrative visualization. We propose a reference model that leverages \rv{foundation} models for crafting narrative visualizations. In particular, we have \rv{identified} four stages of narrative visualization based on previous research, and nine tasks for characterizing the state-of-the-art techniques. With the survey, we connect prior studies by \rv{integrating them into our proposed} reference model. Furthermore, \rv{this} paper explores the relationship between \rv{foundation} models and narrative visualization, highlighting existing challenges and promising directions for future research.  We hope our work could guide researchers in making informed decisions about effectively \rv{incorporating} these advanced models into their work, thereby enhancing the quality and effectiveness of narrative visualization.

\section*{ACKNOWLEDGMENTS}
Nan Cao is the corresponding author. This work was supported in part by NSFC 62072338, 62372327, NSF Shanghai 23ZR1464700, and China Postdoctoral Science Foundation (2023M732674). We would like to thank all the reviewers for their valuable feedback.












\bibliographystyle{unsrt}
\bibliography{template}
\begin{IEEEbiography}[{\includegraphics[width=1in,height=1.25in,clip,keepaspectratio]{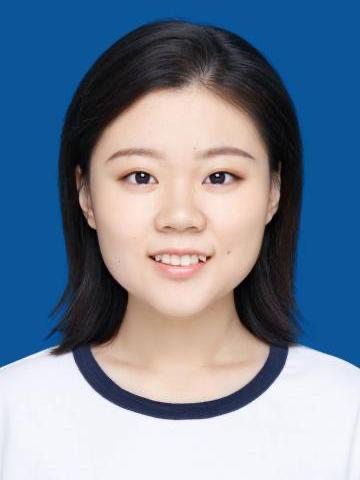}}]{Yi He} received her B.Eng degree from School of Digital Media \& Design Arts, Beijing University of Posts and Telecommunications.  She is currently working toward his Ph.D. degree as part of the Intelligent Big Data Visualization (iDV$^x$) Lab, Tongji
University. Her research interests include data
visualization and artificial intelligence.
\end{IEEEbiography}

\begin{IEEEbiography}[{\includegraphics[width=1in,height=1.25in,clip,keepaspectratio]{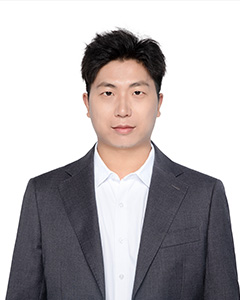}}]{Ke Xu} is currently a tenure-track associate professor at Nanjing University. Prior to this, he was an scientist in the Data Intelligence Lab at HUAWEI Cloud and a postdoctoral research associate in the Visualization and Data Analytics Research Center (VIDA) at New York University (NYU). He obtained his Ph.D. in the Department of Electronic and Computer Engineering at the Hong Kong University of Science and Technology in 2019, and B.S. in Electronic Science and Engineering from Nanjing University, China in 2015. His research interests include data visualization, human-computer interaction, with focus on visual anomaly detection and explainable AI.
\end{IEEEbiography}
\begin{IEEEbiography}[{\includegraphics[width=1in,height=1.25in,clip,keepaspectratio]{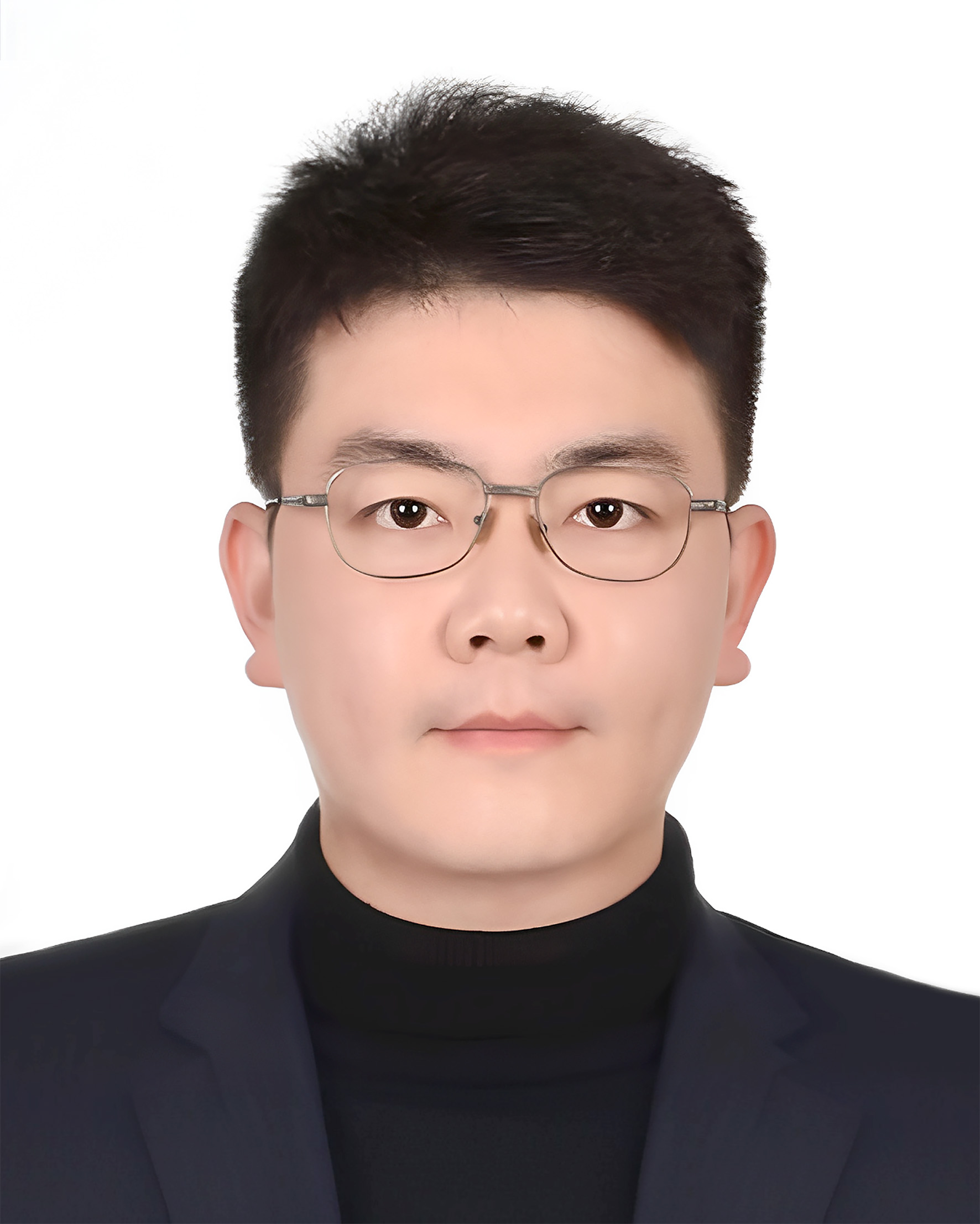}}]{Shixiong Cao} received his Master's degree in Design from Sangmyung University in South Korea in 2019, and subsequently obtained a Ph.D. degree from Sungkyunkwan University in South Korea in 2023. Currently, he works as a postdoctoral researcher at Tongji University, and his research interests include information design, narrative visualization design, and user experience design.
\end{IEEEbiography}
\begin{IEEEbiography}[{\includegraphics[width=1in,height=1.25in,clip,keepaspectratio]{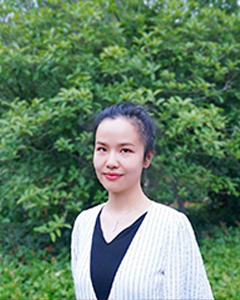}}]{Yang Shi} is an associate professor and doctoral supervisor in the College of Design and Innovation, Tongji University, and a member of the Visualization and Visual Analysis Committee of CSIG. She focuses on the intersection of computer and design, and her research interests are theoretical models for visualization design and intelligent design for visualization. She was awarded the second prize of the Natural Science Award of the Chinese Society of Graphics 2022, and was listed in the 2019 Forbes China 30 Under 30 (Science). Meanwhile, she served as the IEEE VIS2022-2023 Meetup Chair and IEEE PacificVis2020 Poster Chair.
\end{IEEEbiography}
\begin{IEEEbiography}[{\includegraphics[width=1in,height=1.25in,clip,keepaspectratio]{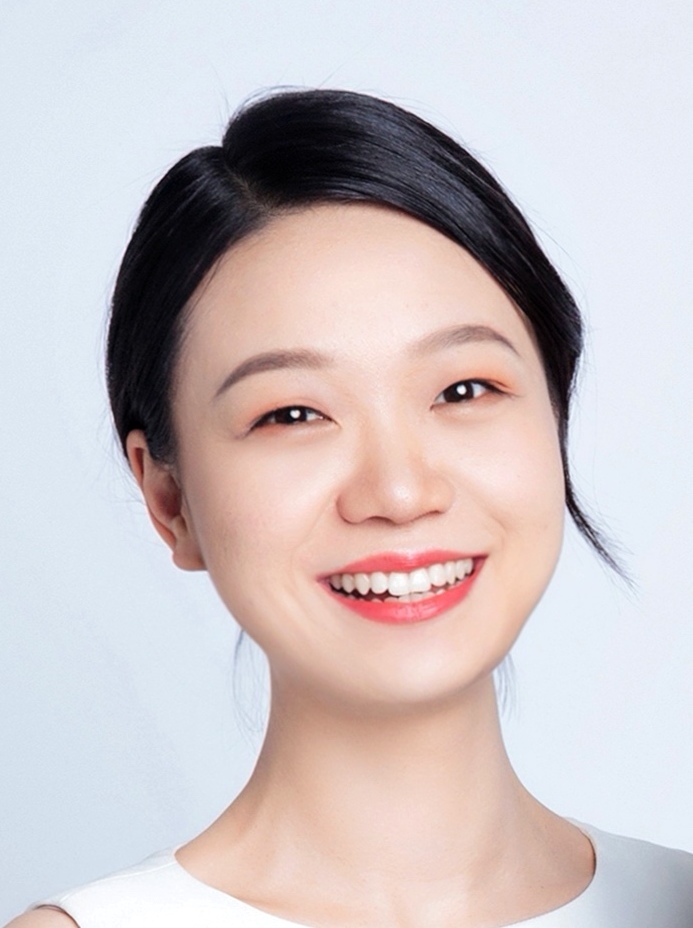}}]{Qing Chen} received her B.Eng degree from the Department of Computer Science, Zhejiang University and her Ph.D. degree from the Department of Computer Science and Engineering, Hong Kong University of Science and Technology (HKUST). After receiving her PhD degree, she worked as a postdoc at Inria and Ecole Polytechnique. She is currently an associate professor at Tongji University. Her research interests include information visualization, visual analytics, human-computer interaction, generative AI and their applications in education, healthcare, design, and business intelligence.
\end{IEEEbiography}


\begin{IEEEbiography}[{\includegraphics[width=1in,height=1.25in,clip,keepaspectratio]{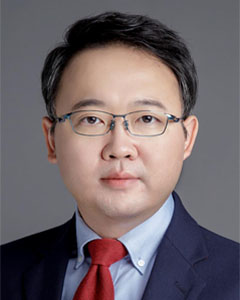}}]{Nan Cao} received his Ph.D. degree in Computer Science and Engineering from the Hong Kong University of Science and Technology (HKUST), Hong Kong, China in 2012. He is currently a professor at Tongji University and the Assistant Dean of the Tongji College of Design and Innovation. He also directs the Tongji Intelligent Big Data Visualization Lab (iDV$^x$ Lab) and conducts interdisciplinary research across multiple fields, including data visualization, human computer interaction, machine learning, and data mining. He was a research staff member at the IBM T.J. Watson Research Center, New York, NY, USA before joining the Tongji faculty in 2016.
\end{IEEEbiography}
\end{document}